\documentclass[a4paper,aps,pra,showpacs,showkeys,preprintnumbers]{revtex4-2}
\usepackage[utf8]{inputenc}
\usepackage[english]{babel}
\usepackage[T2A]{fontenc}
\usepackage{amssymb,amsfonts,amsmath,mathtext,enumerate,float,dsfont}
\usepackage{dblfloatfix}
\usepackage{graphics,graphicx,epsfig,epstopdf}
\usepackage{caption}
\usepackage{cmap}
\usepackage{multirow}
\usepackage{indentfirst}
\usepackage[usenames]{color}
\usepackage{amsthm}
\usepackage{physics}
\usepackage{xcolor}
\usepackage{type1ec}%
\usepackage{dcolumn}%
\usepackage{dblfloatfix}

\begin{document}

\def\BY{\begin{eqnarray}}
\def\EY{\end{eqnarray}}
\def\L{\label}
\def\nn{\nonumber}
\def\ds{\displaystyle}
\def\o{\overline}
\def\({\left (}
\def\){\right )}
\def\[{\left [}
\def\]{\right]}
\def\<{\langle}
\def\>{\rangle}
\def\h{\hat}
\def\td{\tilde}
\def\r{\vec{r}}
\def\ro{\vec{\rho}}
\def\h{\hat}
\def\v{\vec}

\title{Comparison of Controlled-Z operation and beam-splitter transformation for generation of squeezed Fock states by measurement}

\author{E. N. Bashmakova}
\author{S. B. Korolev}
\author{T. Yu. Golubeva } 
\affiliation{St.Petersburg State University, Universitetskaya nab. 7/9, St.Petersburg, 199034, Russia}

\begin{abstract}
In the present paper, we investigate the protocol for the generation of squeezed Fock states by measuring the number of particles from a two-mode entangled Gaussian state using a beam splitter and a controlled-Z operation.  From two different perspectives, we analyzed two entanglement transformations in the protocol. We evaluated the energy costs and resource requirements of the analyzed schemes. Furthermore, we studied the impact of experimental imperfections on the non-Gaussian states generated by measuring the number of particles. We explored the effects of photon loss and imperfect detectors on the squeezed Fock state generation protocol.
\end{abstract}
\maketitle

\section{Introduction}
To date, non-Gaussian states and non-Gaussian operations hold significant importance in quantum computation and quantum information in continuous variables. Several issues motivate this interest.  To build universal quantum computations in continuous-variables, one must be able to implement at least one non-Gaussian operation \cite{Braunstein_2005,Lloyd_1999}. Another significant factor contributing to the increasing interest in non-Gaussian states is their potential application in error correction protocols \cite{Ralph_2003,Hastrup_2022}. Furthermore, the advantages of employing quantum non-Gaussian states as a resource have already been demonstrated in many areas, including quantum metrology \cite{Zhang,Hou,PhysRevResearch.3.033250}, quantum cryptography \cite{Lee2019} and information transmission \cite{Guo2019}. Specifically, non-Gaussian states allow improvements in existing quantum optics and informatics protocols. For instance, in the problem of quantum teleportation in continuous variables, by using auxiliary non-Gaussian states, errors can be significantly reduced \cite{PhysRevA.61.032302,Zinatullin2023,Zinatullin2021}. 

Among non-Gaussian states, the squeezed Fock (SF) states hold a unique position \cite{PhysRevLett.119.033602,PhysRevA.87.062115,Olivares2006,olivares2005squeezed,Kral1990,NIETO1997135,PhysRevA.40.2494,PhysRevA.72.033822,PhysRevLett.105.053602}. It was demonstrated \cite{QECCSFOCK} that Schr\"{o}dinger cat states \cite{Sychev2017,Buzek1995}  and squeezed Schrödinger cat states \cite{bashmakova2023effect,Ourjoumtsev2007} face strong competition from SF states in error-correction protocols. Such states can be used to protect against both photon loss and dephasing errors \cite{Ralph_2003,Hastrup_2022}. In addition, SF states are seen as a valuable resource for quantum metrology \cite{DellAnno}. Moreover, recent research \cite{PhysRevLett.132.230602} has demonstrated that these states serve as an effective resource for generating other beneficial non-Gaussian states, such as the large-amplitude squeezed Schr\"{o}dinger cat and high-quality Gottesman-Kitaev-Preskill states.

Information applications (primarily error correction protocols) operate with idealized states, the generation of which, in most cases, is carried out only approximately \cite{PhysRevLett.132.230602,Podoshvedov_2023,Takase2021}. However, we have proposed a protocol for generating exact (with fidelity equal to 1) SF states. The general consideration of this protocol included one or more photon subtractions from a two-mode entangled Gaussian (TMEG) state. It has been shown that certain conditions must be applied to a TMEG state to generate SF states of an arbitrary order. We have proven that the detection of one photon in one mode of an arbitrary TMEG state should result in the generation of the first SF in the other. For the multiple-photon subtraction, there are conditions that should be imposed on the parameters of the TMEG state in order to generate SF states to be possible.

One can use any entangling operation in the protocol for generating SF states \cite{PhysRevA.109.052428}. However, it has been shown in \cite{korolev2024estimationsetstatesobtained} that protocols with the beam splitter (BS) or the controlled-Z (CZ) gate have the highest probability of generating SF states. In this work, we compare these two experimental setups used for the generation of TMEG states. Based on the theoretical description \cite{PhysRevA.109.052428}, we analyze the impact of entangling operations on the accuracy of the generation of SF states. We show which one of these schemes is more energy- and resource-intensive. We also study the accuracy of the parameters of the TMEG state, which provide the generating SF states.

The paper is structured as follows. Section II addresses the problem of generating SF states by one or more photon subtractions from a non-displaced TMEG state. Section III examines the impact of photon losses and inefficient detectors on the SF state generation protocol. Section IV is devoted to a comparison of the energy costs associated with generating input states through the employment of these two operations. 

\section{Theoretical justification of the possibility of generating squeezed Fock states by measurement}
\label{Gen} 

\begin{figure}[H]
    \centering
    \includegraphics[scale=0.6]{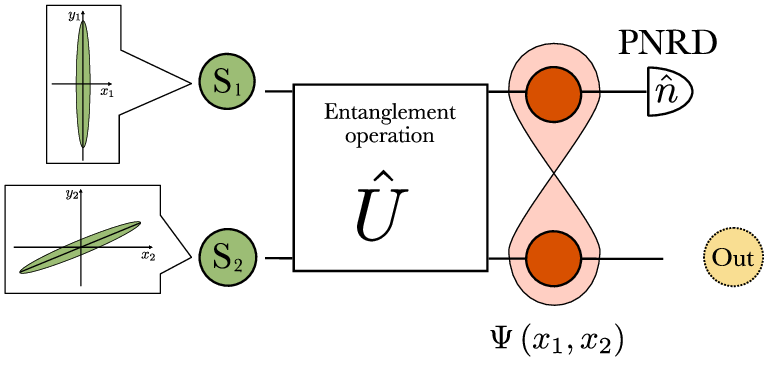}
    \caption{Schemes of the protocol for generating a squeezed Fock state. In the figure: $\text{S}_i$ is the $i$-th squeezed oscillator, $\Psi \left(x_1,x_2\right)$ is the wave function of the output TMEG state, PNRD is photon number resolving detector, and Out is the output reduced state. In  inserts,  the phase portraits  over the quadrature plane are schematically depicted to illustrate the squeezing of the input.}
    \label{fig:generation}
\end{figure}

In this section, we briefly recall the main results obtained in \cite{PhysRevA.109.052428}. We also present the analytical expressions required for further research. We consider the protocol for generating the SF states \cite{PhysRevA.109.052428} in schemes involving the measurement of particle numbers. Fig. \ref{fig:generation} shows a scheme in which two squeezed vacuum states are entangled. One measures the state at one of the outputs using a photon number resolving detector (PNRD).  The unmeasured state at the other output we will call the output state (Out state).

Let us recall the explicit form of the wave function of the $n$-photon SF state for a given squeezing parameter $R$ reads
\begin{align}
\label{SFS}
\Psi_{\mathrm{\mathrm{SF}}}(x, R, n)=\frac{e^{-\frac{1}{2} e^{2 R} x^2} H_n\left(e^R x\right)}{\sqrt{A_n e^{-R}}},
\end{align}
where $A_n=2^n n! \sqrt{\pi}$ and $H_n(x)$ stands for the Hermite polynomials \cite{gradshteyn2014table}.  Here, $R$ is the real squeezing parameter. For brevity, we assume that all squeezing parameters can take both positive and negative values. In this case, a positive and negative squeezing parameter will correspond to the $\hat{x}$-quadrature and $\hat{y}$-quadrature squeezing, respectively.

Using the von Neumann postulate, it has been shown \cite{PhysRevA.109.052428} the wave function of the state in mode 2 (Out state in Fig. \ref{fig:generation}) can be found as follows:
 \begin{align} \label{Psi_out}
\Psi_{out}\left(x, n\right)=
\frac{(-1)^n \sqrt[4]{\mathrm{Re} [a] \mathrm{Re} [d]-\left(\mathrm{Re }[b] \right)^2}}{\sqrt{A_n P_n} }
\sqrt{\frac{2(a-1)^n}{(a+1)^{n+1}}}
e^{-\frac{1}{2} x^2
\left(d-\frac{b^2}{a+1}\right)}
 H_n\left[\frac{b x}{\sqrt{a^2-1}}\right],
\end{align}
where $P_n$ is the probability of measuring $n$ photons in mode 1. Parameters $a,b,d  \in \mathbb{Z}$ are determined by the input states and by the entangling operation and satisfy the following conditions:
\begin{align}
 \label{param_TMSS}
  \mathrm{ Re}[d] >0, \quad   \mathrm{Re}[a]>0, \quad  \mathrm{Re} [a] \mathrm{Re}[d]-\left(\mathrm{Re }[b] \right)^2 >0.
\end{align}

One of the most important results of \cite{PhysRevA.109.052428} is the proof that the "universal solution regime"{} is the conditions guarantees the generation of the $n$-th SF state (\ref{SFS}) with squeezing parameter $R$, when 
\begin{align}
    &d-\frac{b^2}{a+1}=e^{2R}, \label{cond0}\\
    &\frac{b^2}{a^2-1}=e^{2R} \label{cond1}.
\end{align}
\begin{figure}[ht]
    \centering
    \includegraphics[width=1\linewidth]{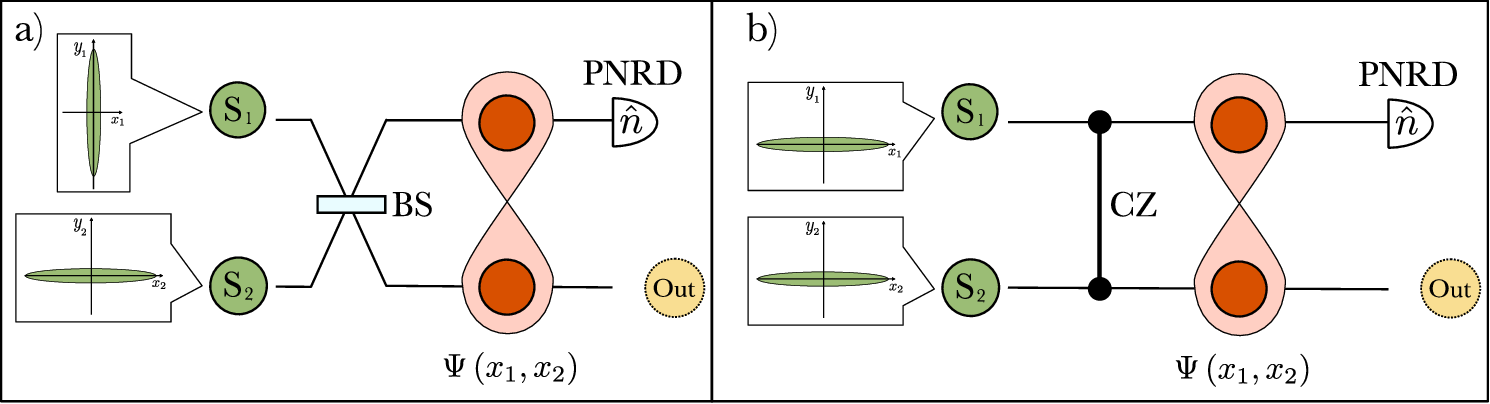}
    \caption{Schemes of the protocol for generating a squeezed Fock state: a) with the beam splitter (with real coefficients $a$, $b$ and $d$); b) with the CZ transformation (with  $a$, $d$ and imaginary $b$). In the figure: $\text{S}_i$ is the $i$-th squeezed oscillator, $\Psi \left(x_1,x_2\right)$ is the wave function of the output TMEG state, PNRD is  the photon number resolving detector, Out is the output reduced state, BS is the beam splitter, and CZ is the CZ transformation. In  inserts,  the phase portraits  over the quadrature plane are schematically depicted to illustrate the squeezing of the input.}
    \label{fig:BS_CZ_gen}
\end{figure}
Therefore, to generate the $n$-th SF state (\ref{SFS}) for a given squeezing parameter  $R$ the conditions (\ref{cond0})-(\ref{cond1}) should be imposed on the parameters of the input states and the entangling operation in the scheme (see Fig. \ref{fig:generation}).

Now let us discuss in more detail the entanglement operation in the protocol for generating quantum non-Gaussian states. As it was shown in \cite{korolev2024estimationsetstatesobtained}, satisfying the conditions of the universal solution regime (\ref{cond0})-(\ref{cond1}) would lead to the requirement that both parameters $a$ and $d$ are real. Divide by the value of $b$ to get the two types of transformations. Real number of $b$ corresponds to BS operation, imaginary number of $b$ corresponds to CZ case. At the same time, according to \cite{korolev2024estimationsetstatesobtained}, these two transformations provide the maximum probability of generating SF states. We will focus on the analysis of these transformations in the protocol under consideration.

Let us start with the BS transformation, described as
\BY 
\hat{U}_{BS}=e^{i\theta (\hat{a}^{\dagger}_1\hat{a}_2+\hat{a}^{\dagger}_2\hat{a}_1)},
\label{BS}
\EY
where $\hat{a}_{i}$ is the photon annihilation operators in the $i$-th input mode. The parameter $\theta$ is determined by the transmission coefficient $t$ as $t=\cos\theta$ (at $\theta\in\[0;\frac{\pi}{2}\]$).

For the BS setup, the degree of squeezing of the two inputs $r_{1}$ and $r_{2}$, and the transmission coefficient of a BS $t$ can be related to the parameters of the output state (\ref{Psi_out}) by the following expressions \cite{PhysRevA.109.052428}:
\begin{align}
   &a= e^{2 r_1} (1-t)+e^{2 r_2}t, \label{param_a}\\
   &b=\sqrt{(1-t) t} \left(e^{2 r_1}-e^{2 r_2}\right), \label{param_b}\\
    &d=e^{2 r_1} t+e^{2 r_2} (1-t). \label{param_d}
\end{align}
We will use these expressions to compare the entangling operations in the protocol for generating SF states.

Next, let us consider the CZ gate for generating SF states:
\begin{align} \label{CZ}
    \Hat{U}_{CZ}=e^{ig \hat{x}_1\hat{x}_2},
\end{align}
where $\hat{x}_i$ is the $x$-quadrature of the $i$-th oscillator and $g$ is the weight coefficient of the CZ gate. To understand what experimental restrictions are imposed on the weight coefficients $g$, let us consider the matrix form of the CZ gate. As is known, the CZ gate with the weight coefficient $g$ transforms the vector of input quadratures into the vector of output quadratures according to the rule
\begin{align}
 \begin{pmatrix}
\hat{x}_{out,1}   \\
\hat{x}_{out,2}  \\
\hat{y}_{out,1}  \\
\hat{y}_{out,2} 
\end{pmatrix} = 
     \begin{pmatrix}
1 & 0 & 0 & 0 \\
0 & 1 & 0 & 0 \\
0 & g & 1 & 0 \\
g & 0 & 0 & 1 
\end{pmatrix}  
 \begin{pmatrix}
\hat{x}_{in,1}   \\
\hat{x}_{in,2}  \\
\hat{y}_{in,1}  \\
\hat{y}_{in,2} 
\end{pmatrix}. 
\label{UCZ}
\end{align}
It was shown in \cite{PhysRevA.106.032414} that using the Bloch-Messiah decomposition \cite{PhysRevA.71.055801} the CZ matrix (\ref{UCZ}) can be represented in the form
\begin{align}
     \begin{pmatrix}
1 & 0 & 0 & 0 \\
0 & 1 & 0 & 0 \\
0 & g & 1 & 0 \\
g & 0 & 0 & 1 
\end{pmatrix} = 
 \begin{pmatrix}
1 & 0 & 0 & 0 \\
0 & 0 & 0 & -1 \\
0 & 0 & 1 & 0 \\
0 & 1 & 0 & 0 
\end{pmatrix}
\begin{pmatrix}
\tau & \rho & 0 & 0 \\
\rho & -\tau & 0 & 0 \\
0 & 0 & \tau & \rho\\
0& 0 & \rho & -\tau
\end{pmatrix}
\begin{pmatrix}
\sqrt{s} & 0 & 0 & 0 \\
0 & \frac{1}{\sqrt{s}} & 0 & 0 \\
0 & 0 & \frac{1}{\sqrt{s}} & 0 \\
0 & 0 & 0 & \sqrt{s} 
\end{pmatrix}
\begin{pmatrix}
\rho & \tau & 0 & 0 \\
\tau & -\rho & 0 & 0 \\
0 & 0 & \rho & \tau \\
0 & 0 & \tau & -\rho 
\end{pmatrix}
\begin{pmatrix}
1 & 0 & 0 & 0 \\
0 & 0 & 0 & 1 \\
0 & 0 & 1 & 0 \\
0 & -1 & 0 & 0 
\end{pmatrix}, 
\label{UCZBM}
\end{align}
where
\begin{align}
&\rho=\frac{\sqrt{s}}{\sqrt{1+s}},\\
&\tau=\frac{1}{\sqrt{1+s}}, \\
&s=\frac{1}{2}(2+g^{2}-g\sqrt{4+g^2}).\label{sg}
\end{align}
Here, the first and last matrices describe the phase shifters. The second and fourth matrices describe the BS transformation with transmittance $\tau$ and reflection coefficient $\rho$, respectively. The third matrix corresponds to the squeezing with strength $s$. Since we consider the case of non-negative weight coefficients $g$, $s\in [0,1]$. From the decomposition (\ref{UCZBM}), it is clearly seen that the CZ gate contains a squeezing operation. More precisely, implementing the CZ gate involves two auxiliary squeezed oscillators \cite{PhysRevLett.101.250501}. 
Strictly speaking, to achieve an exact CZ gate implementation, auxiliary oscillators must be squeezed perfectly. However, one can consider an approximate operation of CZ. Than it is possible to specify a requirement for squeezing auxiliary oscillators, such that the error in the execution of the CZ gate is small compared to the variance of the input oscillators. According to \cite{PhysRevA.106.032414}, one can estimate the squeezing of the auxiliary oscillators $r_{an}$ by the expression (\ref{sg}):
\BY
e^{2r_{an}}\gg\frac{1}{2}(2+g^{2}-g\sqrt{4+g^2}).\label{rg}
\EY

There is currently no experimental design that allows the CZ gate to be implemented exactly for all $g$ values.

For the CZ setup (Fig. \ref{fig:BS_CZ_gen}b), the explicit relationship between the parameters of the input state and the output state for generating the SF state follows form \cite{PhysRevA.109.052428}:
\begin{align}
    &a=e^{2r_1},  \label{CZ1}\\
    &b=ig, \label{CZ2}\\
    &d=e^{2r_2}. \label{CZ3}
\end{align}

We conclude that two common experimental setups can be done by using for the efficient generation of SF states. This could be the CZ gate or the BS transformation. In the next section, we will examine the sensitivity of each scheme to photon losses and detection imperfections.

\section{The effects of photon loss and imperfect detectors in particle number measurement schemes}\label{losses}
\subsection{Uniform losses in particle number measurement schemes}
As in the previous section, we consider the scheme for generating the SF states proposed in \cite{PhysRevA.109.052428}. Nevertheless, \cite{PhysRevA.109.052428} did not address the influence of experimental imperfections on the generated non-Gaussian states, which are inevitably contained in the quantum information protocol. In this section, we investigate the impact of losses on the non-Gaussian state generation protocol. In the subsection \ref{error_PNRD}, we discuss the impact of an imperfect PNRD on generating SF states.

\begin{figure}[H]
    \centering
    \includegraphics[scale=0.5]{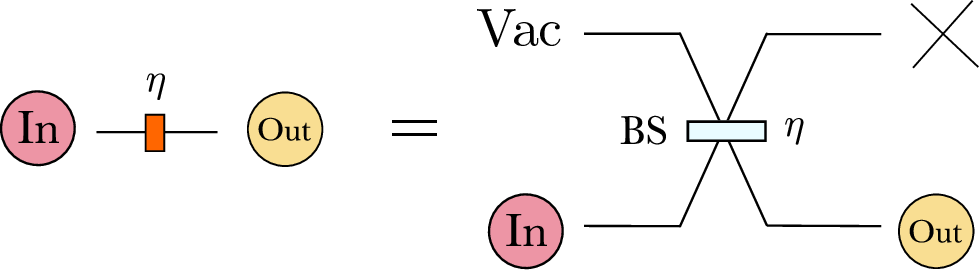}
    \caption{The beam splitter model of small uniform losses. On the left, we have the mode passing through a lossy element with the quantity of loss $\eta$. On the right, we have an equivalent scheme where the lossy element is replaced by a beam splitter with a transmission coefficient $\eta$. In the figure: In is the input state,  Out is the output state, Vac is the vacuum mode, and BS is the beam splitter. A cross indicates that the corresponding channel is not examined.}
    \label{fig:losses_block}
\end{figure}

\begin{figure}[H]
    \centering
    \includegraphics[scale=0.18]{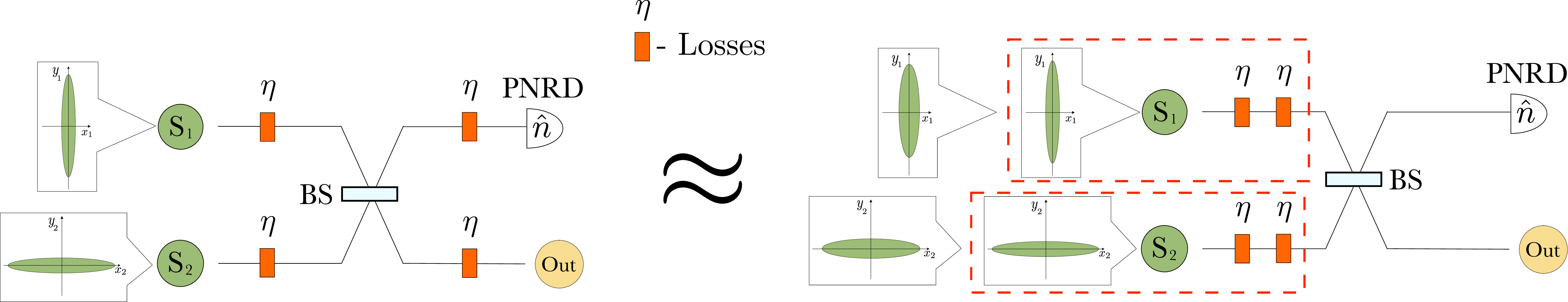}
    \caption{Schemes for generating a squeezed Fock state given small uniform losses $\eta$. Orange rectangles indicate the losses. It is shown that uniform losses commute with passive linear optical operations in the same mode. It is shown that small uniform losses are reduced to decrease of squeezing of the input states. In the figure: $\text{S}_i$ is the $i$-th squeezed oscillator, PNRD is  photon number resolving detector, Out is the output reduced state. In inserts, the phase portraits over the quadrature plane are schematically depicted to illustrate the squeezing of the input.}
    \label{fig:shems_losses}
\end{figure}

We consider the beam splitter loss model \cite{DEMKOWICZDOBRZANSKI2015345,PhysRevA.57.2134}. It is the most common way of modelling losses in quantum optics. In this model, we have the mode passing through a lossy element with the quantity of loss $\eta$. It is equivalent to replacing the loss by a beam splitter with transmittance $\eta$, as in Fig. \ref{fig:losses_block}. We deal with the uniform losses case, where all modes have the same loss. It should also be noted that modern optical materials obey the high experimental requirements, i.e. $\eta\ll 1$.

Losses can occur at every step of the optical scheme. Furthermore, losses can occur both before and after the entangling operation. However, it was shown in \cite{Oszmaniec_2018} that small uniform losses commute with passive linear optical operations in the same mode.
This idea is effective when discussing the setup with a beam splitter.
In this case all small losses can be transferred to the input of the system (see Fig. \ref{fig:shems_losses}). Next losses can be reduced to the decrease of squeezing parameters of the input states \cite{PhysRevLett.124.100502,GarciaPatron2019simulatingboson,MartinezCifuentes2023classicalmodelsmay}. In Fig. \ref{fig:shems_losses}, it is schematically depicted by a noise ellipse of a larger width.

\begin{figure}[t]
   \begin{center}
    \includegraphics[scale=0.5]{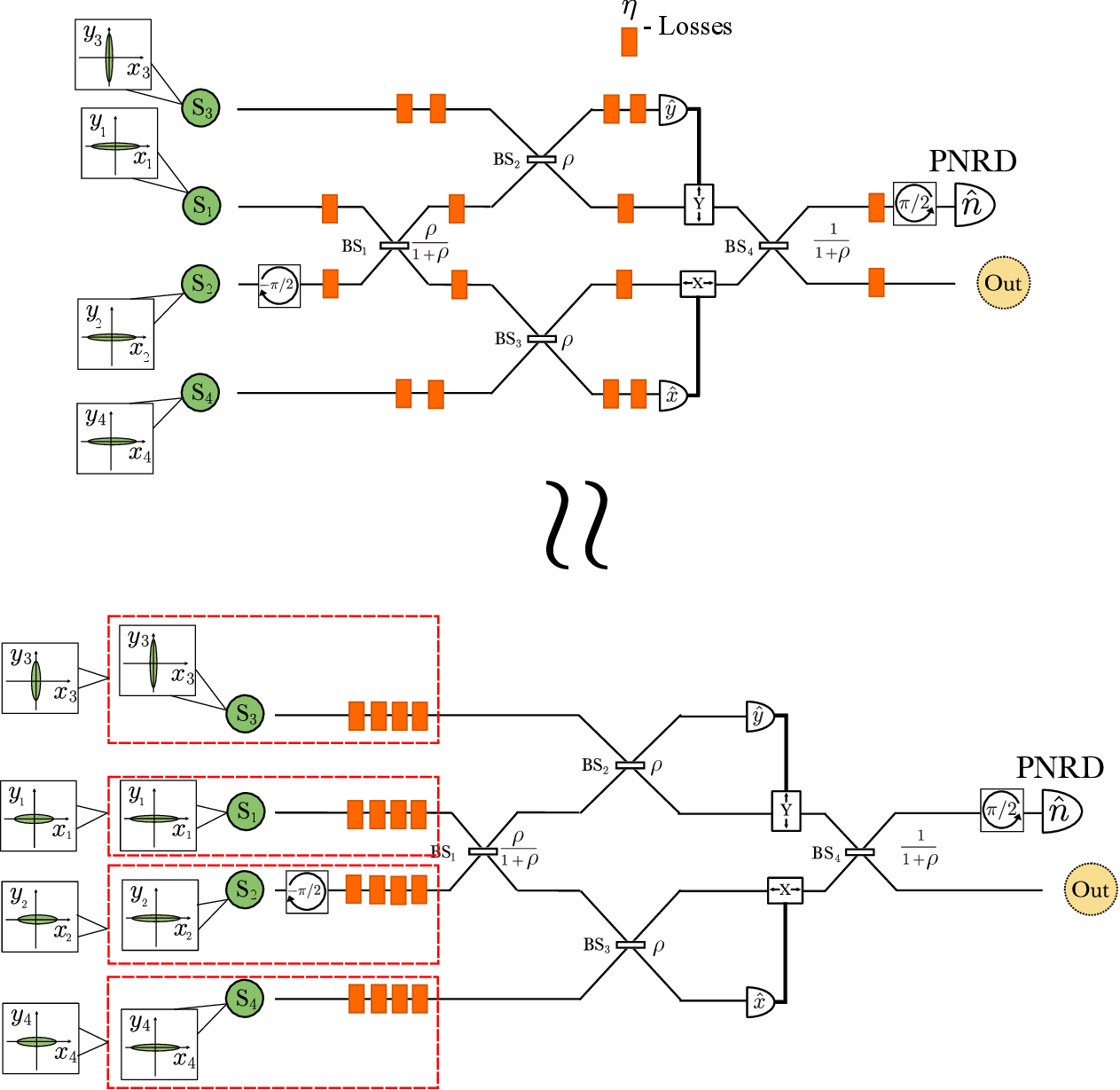}
    \caption{Schemes for generating a squeezed Fock states given small uniform losses $\eta$ considering the optical implementation of the CZ gate. Orange rectangles indicate the losses. It is shown that uniform losses commute with passive linear optical operations in the same mode. In the figure: $S_1$, $S_2$, $S_3$ and $S_4$ are two squeezed oscillators and two squeezed auxiliary oscillators,respectively; PNRD is  photon number resolving detector; Out is the output reduced state. In inserts, the phase portraits over the quadrature plane are schematically depicted to illustrate the squeezing of the input.}
    \label{fig:CZ_err}
    \end{center}
\end{figure}

However, under the general consideration, it is not possible to perform a similar loss transfer in a CZ setup since it is an active operation. Limiting ourselves to the optical implementation of the CZ gate discussed in Section II, we can represent the CZ operation as a set of passive operations and auxiliary squeezed oscillators. Then a similar transfer of losses would also affect the squeezing of the input oscillators, as well as the auxiliary ones. This is illustrated schematically in Fig. \ref{fig:CZ_err}.
 Obviously, changing the squeezing parameters of input states takes us away from the exact satisfaction of the universal solution regime. 
Furthermore, the squeezing of the auxiliary oscillators will become even worse when losses are transferred to the CZ gate input, and the transformation will perform worse.
 It is necessary to answer the following questions. How close are the states generated in lossy channel to the required ones? Does the type of entangling operation affect the tolerance of the protocol to losses? Next, we will answer these questions.

The universal solution regime (\ref{cond0})-(\ref{cond1}) allows us to connect the squeezing of input oscillators to each parameter of the output state. One can choose input parameters so that the highest probability of the SF generation occurs. Thus, for each SF state with number $n$ and squeezing degree $R$, there are a transmittance coefficient $t$ of the BS setup and a weight coefficient $g$ of the CZ case. These parameters of the entangling operation provide the maximum probability of generating the corresponding quantum state. Hereinafter, we will call these values $t$ and $g$ as optimal. Thus, for each SF state, we can associate three parameters: two values of the squeezing input oscillators and one parameter of the entangling operation.

Discussing the BS setup, existing losses we interpret as a probabilistic process of generating input squeezed states not with fixed required values of $r_{1}$ and $r_{2}$, but with values lying in the range $[0, \;r_{opt_{i}}]$ (for $i=1,2$). Such limits conditioned by the fact that losses can only reduce the squeezing degree of oscillators. The lowest degree of squeezing equals zero, which corresponds to the vacuum state. This process is described by the Gaussian distribution of squeezing parameters in each mode as
\BY
\rho_{i}(r_{i}, \mu)=\frac{2}{\sqrt{2\pi} \erf(\frac{1}{\sqrt{2}\mu})}\frac{1}{ |\mu \;r_{opt_{i}}|} &\exp\Big(-\frac{1}{2}\Big(\frac{r_{i}-r_{opt_{i}}}{|\mu \;r_{opt_{i}}|}\Big)^{2}\Big), \quad i=1,2.
\label{distrib}
\EY
Here, $r_{opt_{i}}$ is the squeezing parameter of the $i$-th mode according to the universal solution regime (\ref{cond0})-(\ref{cond1}). It is important to note that this distribution density is normalized on the interval $[0, \;r_{opt_{i}}]$. Deviation $\mu$ of the squeezing parameter from $r_{opt_{i}}$ does not lead to an increase in squeezing. Therefore, the quantity of loss $\eta$ is determined by the deviation $\mu$. Note that, before discussing losses, we considered pure input states. When transferring losses to the input of the scheme, we are already dealing with mixed input states. These input states are a mixture of oscillators with different squeezing degrees. This is shown in the distribution (\ref{distrib}).

 When discussing the CZ setup, we must similarly consider the effect of losses on the auxiliary oscillators, in addition to the process described by the expression (\ref{distrib}). We will consider the maximum achievable squeezing of ancillaries before losses to be -15 dB \cite{PhysRevLett.117.110801}, which corresponds to $r_{an}^{max}=1.7$. Then the effect of losses can be considered by the distribution $\rho_{i}^{an}(r_{i}^{an}, \mu)$, similar to (\ref{distrib}), where $r_i$ is replaced by $r_i^{an}$, and $r_{opt_i}$ should be replaced by $r_{an}^{max}$. Thus, in this case, losses are reduced not only to the degradation of optimal SF generation parameters, but also to the degradation of the CZ operation.

Let us investigate how losses affect the SF states generated. This analysis will examine degradation of the probability and the fidelity of the generating states under consideration. Using the distribution (\ref{distrib}), the average probability $\overline{P}_{n}(\mu)$ and the average fidelity $\overline{F}_{n}(\mu)$ of generating the required SF state can be defined as:
\BY
&\overline{P}_{n}(\mu)=\int dr_{1} \;dr_{2}\;\rho_{1}(r_{1}, \mu)\rho_{2}(r_{2}, \mu)P_{n}(r_{1}, r_{2}, t)\label{p_BS},\;\;\;
\\& \overline{F}_{n}(\mu)=\int dr_{1} \;dr_{2}\;\rho_{1}(r_{1}, \mu)\rho_{2}(r_{2}, \mu) F_{n}(r_{1}, r_{2}, t)\;\;\;\label{F_BS}
\EY
-- for the BS setup, and
\BY
&&\overline{P}_{n}(\mu)=\int dr_{1} \;dr_{2}\;dr_1^{an}\;dr_2^{an}\;\rho_{1}(r_{1}, \mu)\rho_{2}(r_{2}, \mu)\;\rho_{1}^{an}(r_{1}^{an}, \mu)\;\rho_{2}^{an}(r_{2}^{an}, \mu)\;P_{n}(r_{1}, r_{2}, r_{1}^{an}, r_{2}^{an}, g)\label{p_CZ},
\\&& \overline{F}_{n}(\mu)=\int dr_{1} \;dr_{2}\;dr_1^{an}\;dr_2^{an}\;\;\rho_{1}(r_{1}, \mu)\rho_{2}(r_{2}, \mu)\;\rho_{1}^{an}(r_{1}^{an}, \mu)\;\rho_{2}^{an}(r_{2}^{an}, \mu)\; F_{n}(r_{1}, r_{2}, r_{1}^{an}, r_{2}^{an}, g) \label{F_CZ}
\EY
-- for the CZ one.

Here, $P_{n}(r_{1}, r_{2}, t)$ and $P_{n}(r_{1}, r_{2}, r_{1}^{an}, r_{2}^{an}, g)$ are the probabilities, and $F_{n}(r_{1}, r_{2}, t)$ and $F_{n}(r_{1}, r_{2}, r_{1}^{an}, r_{2}^{an}, g)$ are the fidelities of generating the $n$-th SF state. Transmittance coefficient $t$ and weight coefficient $g$ have chosen for generating the SF state with maximum probability. We assume that the input squeezed states are statistically independent by introducing two different Gaussian distributions for the squeezing degree of the input states.
In Eqs. (\ref{p_BS})-(\ref{F_CZ}), the fidelity is defined as
\begin{align}  \label{fidelity}
 F_n=\langle \Psi_{\mathrm{\mathrm{SF}}}(n,R) |\hat{\rho}_{out}|\Psi_{\mathrm{\mathrm{SF}}}(n,R)\rangle{},
 \end{align}
and the probability gets the following form:
\begin{align}  \label{probability}
P_{n}=Sp(\hat{\rho}_{out}).
\end{align}
Here $\Psi_{\mathrm{\mathrm{SF}}}(n,R)$ is the wave function of the squeezed Fock state with a certain number $n$ and the squeezing degree $R$. Arguments of the $F_n$, as well as the  density matrices of the output state $\hat{\rho}_{out}$, are determined by the parameters of the scheme.
Note that, for each SF state with the required squeezing parameters in the expressions (\ref{p_BS})-(\ref{F_CZ}), we used the limits of integration such that integration was carried out over a physically determined range of squeezing degree. For positive values $r_{i}$, it is the interval $[0, \;r_{opt_{i}}]$, and for negative values $r_{i}$, it is $[r_{opt_{i}}\;0]$.

\begin{figure}[t]
    \centering
    \includegraphics[scale=0.55]{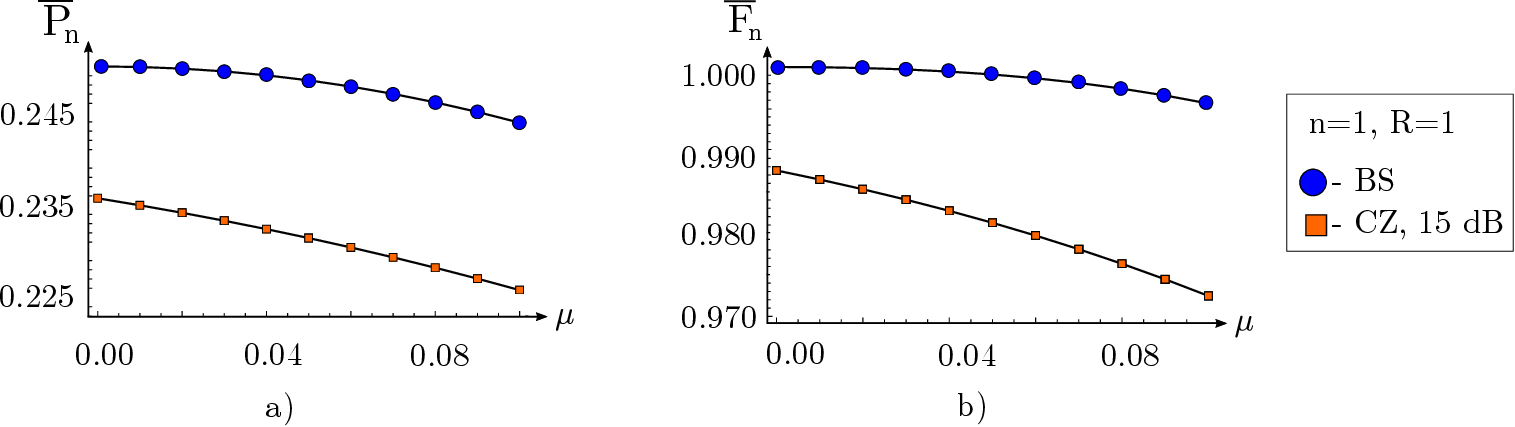}
    \caption{Dependence of the integral characteristics of the first squeezed Fock state at the squeezing degree $R=1$ on the effective quantity of loss $\mu$ using a beam splitter transformation (marked with a blue circle data marker) and imperfect CZ gate with fidelity 0.99 (marked with a square orange data marker): a) the average probability; b) the average fidelity. In the calculations for the squeezed Fock state, the optimal transmittance $t$ of the beam splitter and the weight coefficient $g$ for the CZ gate were used.}
\label{ris:sqloseesfin}
\end{figure}
As an example, let us calculate the average probability and average fidelity of generating a certain SF state by fixing the values of $\mu$. We consider the first SF state at a fixed output $R=1$. For calculating the integral characteristics of the considered state, the optimal transmittance $t$ of the beam splitter and the weight coefficient $g$ for the CZ gate were used. The results are illustrated in Fig. \ref{ris:sqloseesfin}. It was shown in \cite{PhysRevA.109.052428} that the maximum probability of generating the first SF state is 25\% in both BS and ideal CZ setups. However, we are performing calculations for realistically achievable parameters here. So we limit the squeezing degree of the ancillary oscillators in the CZ setup to -15 dB. This results in a slight degradation in both probability and fidelity, even without considering photon losses.  In Fig. \ref{ris:sqloseesfin}a, the deviation of the average probability from its maximum (25\%) of generating the first SF state is shown. Even with an effective loss $\mu=10\%$, the average probability at BS setups decreases by less than 1\%. It can be seen from Fig. \ref{ris:sqloseesfin}a that the CZ gate protocol started from lower probability (because of its imperfection) and has a higher decrease in probability induced by photon losses than the BS case. Similar results are observed for other values of $R$ and for the generation of SF states with another numbers of $n$.

We observe a similar picture also regarding the average fidelity of the two schemes
(Fig. \ref{ris:sqloseesfin}b). For BS setup, the average fidelity is robust to losses. With an effective quantity of loss of 10\% at a BS setup, the average fidelity does not fall below the level of 0.99. In the CZ case, the fidelity is degraded both by the imperfect implementation of the gate and by photon losses. However, even in this case, the fidelity drop is below 3\% for optical losses of 10\%. Thus, it can be argued that the BS setup has the advantage of being more robust in its integral characteristics to losses. Furthermore, we observe that this behavior of the integral characteristic is also preserved for SF states with a different $R$ and other numbers $n$.

We have demonstrated the stability of the SF state generation to losses. However, we would like to explore the range of values of the squeezing parameters of input oscillators that provide fidelity equal to 0.99.

\begin{figure}[t]
    \centering
    \includegraphics[scale=0.45]{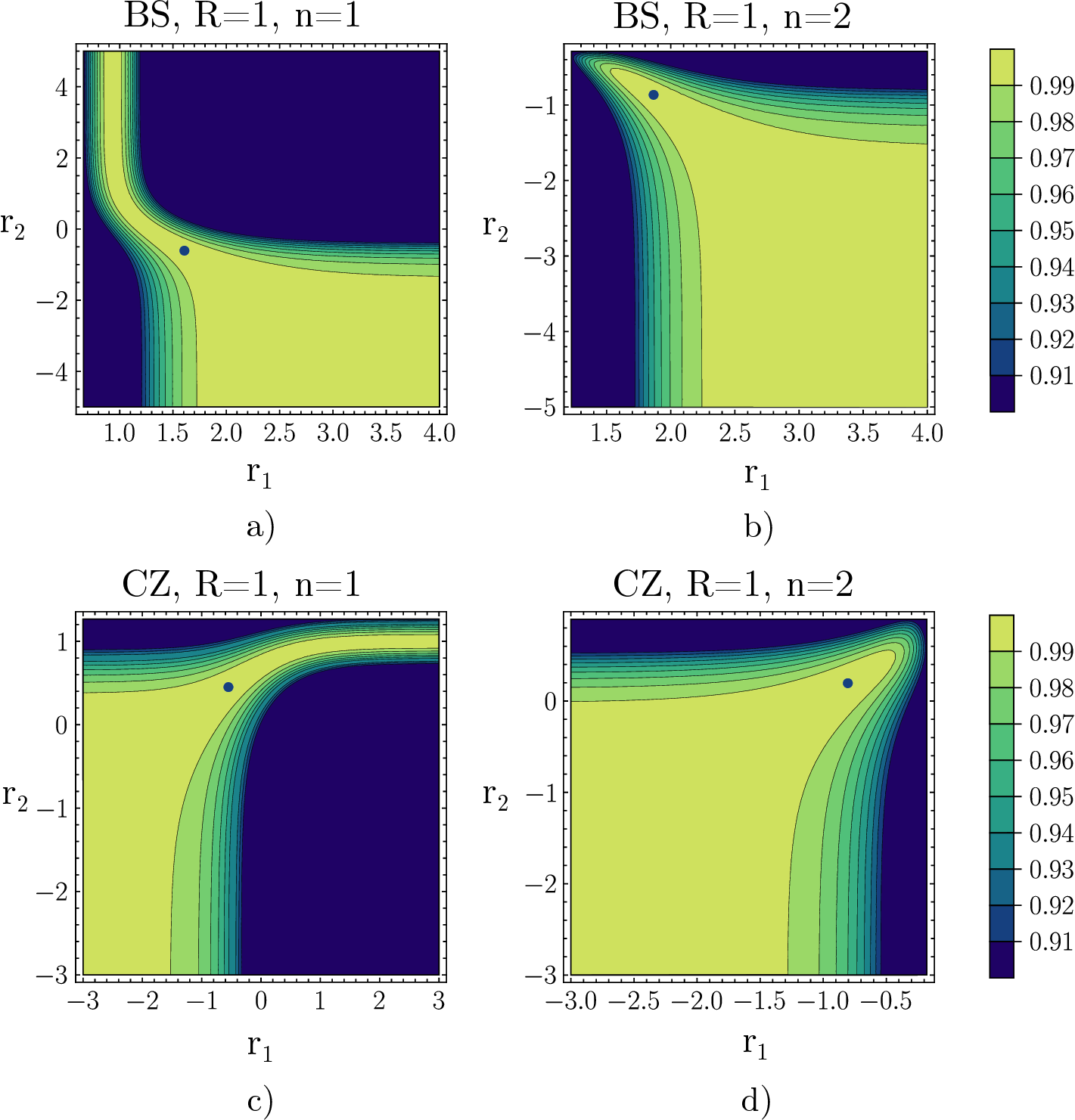}
    \caption{Dependence of the fidelity of generating the squeezed Fock states on the squeezing degree of input states $r_{1}$ and $r_{2}$ at a fixed output squeezing parameter $R=1$ and state number $n$ in: a beam splitter setup with a) $ n=1$; b) $n=2$; CZ setup with c) $n=1$; d) $n=2$. In these calculations, for each considered squeezed Fock state with number $n$ and squeezing degree $R=1$, the optimal transmittance $t$ of the beam splitter and the weight coefficient $g$ for the CZ gate were used. The dot on each graph marks the squeezing degree of the input oscillators, corresponding to the universal solution regime (\ref{cond0})-(\ref{cond1}).}
\label{fig:error_otn}
\end{figure}

To answer this question, we analyze the fidelity of generating the first and second SF states at a fixed $R=1$. As before, in all calculations, we will use the optimal transmittance $t$ of the BS and the weight coefficient $g$ of the CZ gate. Fig. \ref{fig:error_otn} presents the dependence of the fidelity of generation of the first and second SF states on the input squeezing parameters $r_{1}$ and $r_{2}$ using two entangling operations. On the fidelity surfaces, a dot marks the squeezing degree of the input oscillators corresponding to the universal solution regime (\ref{cond0})-(\ref{cond1}). One can see that in each of the frames, there is a plateau of values $r_{1}$ and $r_{2}$, providing values of fidelity above 0.99 (marked in light green). Fig. \ref{fig:error_otn} shows that in both setups, even without “exactly hitting” {} requiring squeezing parameters of the input oscillators, generating the SF states can be provided with a fidelity level of 0.99.

Comparing the fidelity surfaces of generating SF states in the setups under consideration, one can see a general pattern of fidelity behavior that is observed. However, for a fixed number $n$ and squeezing degree $R$ of the SF state, the range of $r_{1}$ and $r_{2}$, at which the fidelity of 0.99 is ensured, is larger in the CZ setup than in the BS case. The range of input squeezing parameters corresponding to the fidelity level of 0.99 decreases as the number $n$ of the state grows.

In the graphs shown in Fig. \ref{fig:error_otn} one can also notice the asymmetrical dependence of the fidelity on the $r_{1}$ and $r_{2}$ squeezing input parameters. Remember here we are considering the procedure for generating the first and second SF states with the degree of squeezing $R=1$ in $y$-quadrature. Consequently, the output states are asymmetrical on the phase plane regarding the rotation of the quadratures’ plane. This asymmetry of states is the cause of the asymmetry in plots. In addition, one can observe that fidelity has a region where it is the least sensitive to changes in the squeezing parameters of the input oscillators. This area corresponds to a higher squeezing degree of the input oscillators.

These results show that generating the SF states by the two entangling operations is quite resistant to low losses in terms of the stability of the average probability and average fidelity. The BS setup is more resistant to losses than the CZ case. This is explained by the fact that the CZ gate itself contains a squeezing operation which suffer from the losses.

\subsection{The impact of an imperfect detector on the protocol for generating squeezed Fock states}\label{error_PNRD}

PNRD (photon number resolving detector) is an important class of detectors for quantum optics and quantum information. Each detector is characterized by such a parameter as the quantum efficiency $\eta_{eff}$. The quantum efficiency (QE) $\eta_{eff}$ of the detector is determined by comparing the number of incident photons that cause a useful signal to the total number of incident photons on the detector. In the work \cite{PhysRevA.109.052428} it was assumed that $\eta_{eff}=1$. However, in real experiments, PNRD can have a rather high QE, but not an ideal one, i.e. $\eta_{eff}\textless1$. Such QE values can lead to a deterioration of the characteristics of the output states. In this section, we evaluate the impact of the QE of the PNRD on the output states in the considered non-Gaussian state generation protocol in the lossless scheme under the conditions of the universal solution regime (\ref{cond0})-(\ref{cond1}).

\begin{figure}[t]
    \centering
    \includegraphics[scale=0.6]{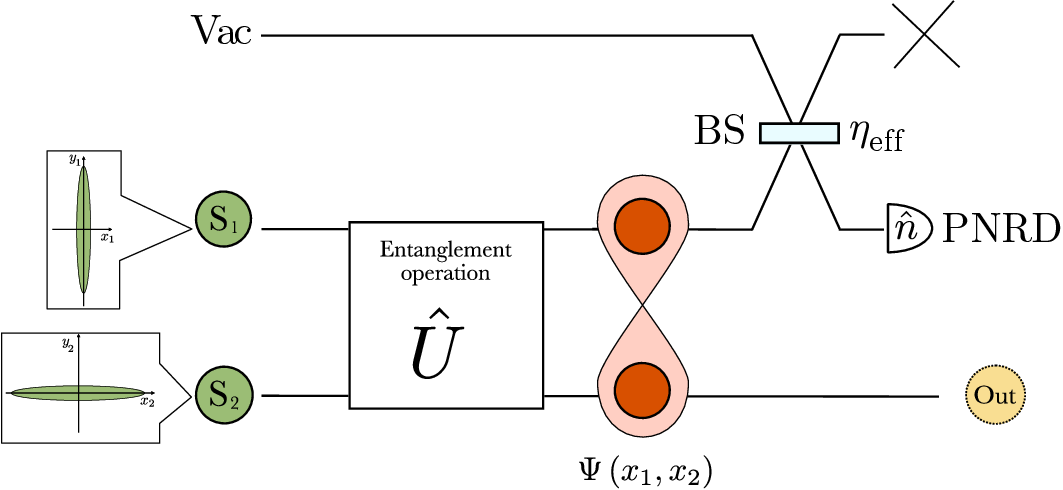}
    \caption{PNRD with quantum efficiency $\eta_{eff} \textless1$ in a scheme for generating squeezed Fock states. Considering the imperfect PNRD occurs by introducing an additional beam splitter (BS), which mixes the measured mode with the vacuum one. The quantum efficiency of the detector $\eta_{eff}$ corresponds to the energy reflection coefficient of such a beam splitter. In the figure: $\text{S}_i$ is the $i$-th squeezed oscillator, Vac is a vacuum mode, $\Psi \left(x_1,x_2\right)$ is the wave function of the output TMEG state, PNRD is  photon number resolving detector, Out is the output reduced state. In  inserts,  the phase portraits  over the quadrature plane are schematically depicted to illustrate the squeezing of the input. A cross indicates that the corresponding channel is not examined.}
    \label{fig:PNDR}
\end{figure}

The imperfect detector occurs by introducing an additional beam splitter that mixes the measured mode with the vacuum one (Fig. \ref{fig:PNDR}). Then the QE of the detector $\eta_{eff}$ corresponds to the energy reflection coefficient of such a beam splitter \cite{PhysRevA.99.032302}.

The fidelity of the generated state has the form (\ref{fidelity}) \cite{Jozsa_fidelity}.
 Let us estimate the influence of the QE $\eta_{eff} \textless1$ of the PNRD on generating the non-Gaussian states. For this calculation, we investigate the dependence of the fidelity of generation of the $n$-th SF state on the QE of the detector $\eta_{eff}$, as
 \begin{align}  \label{fidelity_ffin}
 F_{n}(\eta_{eff})=\int dx_{3}\left|\int dx_{2}\Psi_{out}(x_2, x_3, n,\eta_{eff})\Psi_{\mathrm{\mathrm{SF}}}(x_{2},R, n)\right|^{2},
 \end{align}
 where integration over the $x_{3}$ variable corresponds to partially tracing out the vacuum subsystem. The derivation of this formula is shown in Appendix \ref{PNRD}.

Let us study the dependence of the fidelity of the first and second SF states generating at a fixed squeezing degree of $R=0.5$ and $R=1$ on the QE  of the detector in the BS and CZ setups. We would like to analyze which of the two types of entangling operations has an advantage in terms of protocol resistance to the value of detection QE.

\begin{figure}[t]
   \begin{center}
    \includegraphics[scale=0.4]{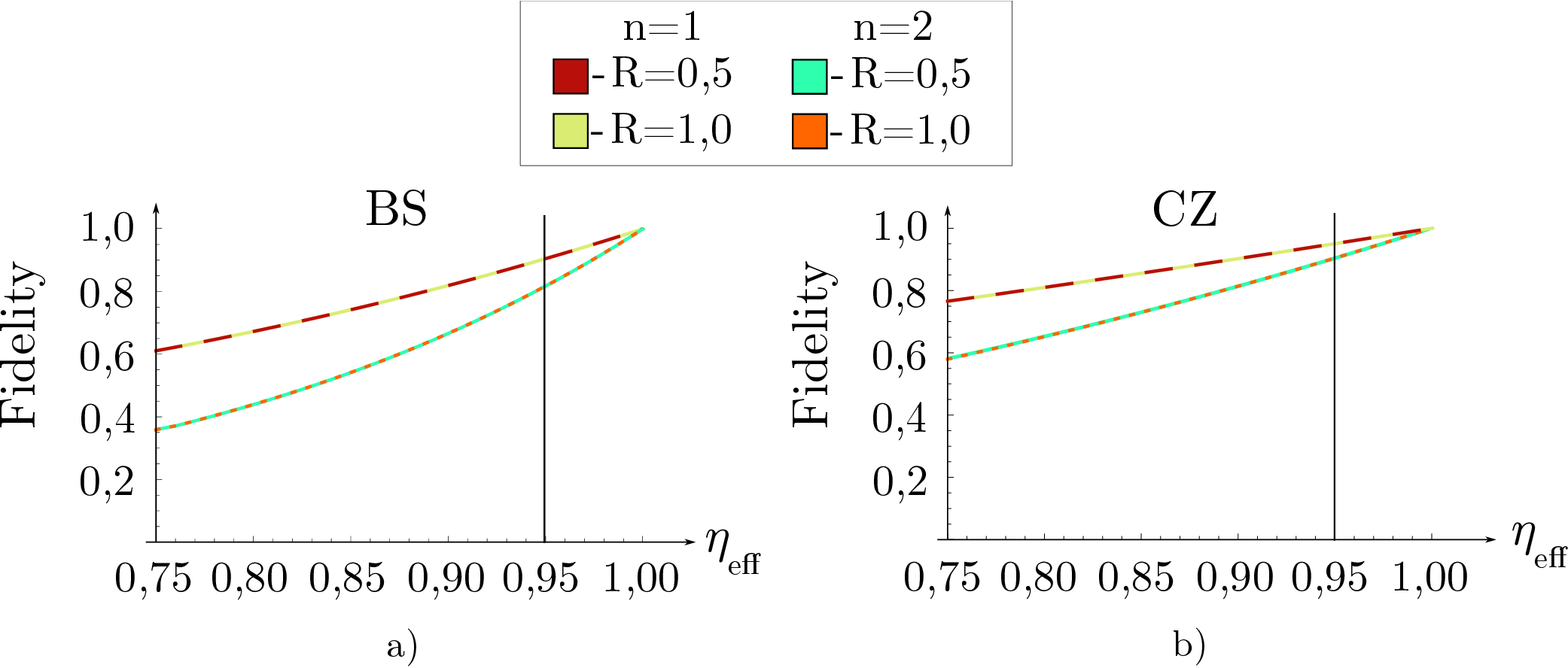}
    \caption{Dependence of the fidelity of generating the first and second squeezed Fock states on the quantum efficiency of the detector $\eta_{eff}$ at a fixed squeezing degree of the output states $R=0.5$ and $R=1$ in: a) a beam splitter setup; b) a CZ setup. The $x$-axis  shows the quantum efficiency of detectors $\eta_{eff}$. The vertical line marks the maximum of the quantum efficiency $\eta_{eff}=0.95$, accessible to modern experiments. In these calculations in the lossless mode, for each considered squeezed Fock state with number $n$ and squeezing degree $R$, the requirement squeezing input parameters of $r_{1}$ and $r_{2}$ and the optimal transmission coefficient $t$ (for the BS setup) and weight coefficient $g$ (for the CZ setup) were used.}
    \label{fig:PNDR_err}
    \end{center}
\end{figure}
 
Fig. \ref{fig:PNDR_err} shows these results. The $x$-axis represents the QE of detectors typical for in real experiments. As one would expect, the maximum fidelity is achieved at the maximum reflection coefficient $\eta_{eff}=1$, when all information about the state remains in the measured mode.
 
The dependence of the fidelity on the QE of the detector highlights an interesting point. One can see that for the two generation protocols (the BS (Fig. \ref{fig:PNDR_err}a) and the CZ (Fig. \ref{fig:PNDR_err}b) setups) the fidelities of the SF states are independent of the squeezing degree of the generated state and is determined only by the state number $n$. Using the PNRD with $\eta_{eff} \textless1$ in the protocol for generating SF states can lead to incorrect interpretation of measurement results that means incorrect number $n$. It can be seen more clearly by considering this issue in the Fock representation (see Appendix \ref{POVM}). 

Let us compare the dependence of the fidelity on the QE of the PNRD for the two transformations. One could observe their general behavior. Fig. \ref{fig:PNDR_err} demonstrates that as the state number $n$ increases, the decay of the fidelity increases as well. However, for the CZ setup, the resulting curves are located higher than in the BS case. In Fig. \ref{fig:PNDR_err}, the vertical line marks the level of QE of the detector, which is equal to 0.95. This value corresponds to the maximum of the quantum efficiency of the detector accessible to modern experiments \cite{Lita:08,10.1063/5.0204340}. In the protocol for generating SF states using the CZ gate, at a given value of the quantum detection efficiency, the fidelity for the first SF state is 0.95, and for the second is 0.9, respectively. Using a BS with QE of the PNRD $\eta_{eff}=0.95$, the fidelity for the first SF state is 0.91, and for the second is 0.82, respectively. It follows from this that the protocol for generating SF states using the CZ gate is more resistant to the QE of detectors. Thus, the protocol for generating SF states using the CZ gate has here advantage over the BS case.

These estimates are correct for the ideal CZ gate with perfectly squeezed auxiliary oscillators. It should be noted that using ancillaries with a finite squeezing degree will degrade the generation of SF states at any level of loss. However, as we can see from the results of the previous section, this degradation is not as critical as the effect of imperfect detection. So, the resistance to imperfect detection should lead us to state the advantages of the CZ setup over the BS one.

\section{Analysis of the energy costs of the two types of entangling transformation: a beam splitter and a CZ gate}\label{energy}

In the previous section, we examined fault-tolerance of the method for generating SF states based on the measurement of particle numbers. We have shown that the BS transformation and CZ gate are competitive in terms of photon loss and measurement imperfections. In this section, we want to compare these two entangling operations to generate the non-Gaussian states basing on an analysis of the energy costs and the limit of input squeezing required for SF generation.

Note that generating a non-Gaussian state can be divided into several stages: generating squeezed input states, performing a unitary entangling transformation, and measuring the number of particles in one of the output modes. It is well-known that the measurement of the number of photons is an energy cost-free operation. The energy costs for generating the SF state (Fig. \ref{fig:BS_CZ_gen}) are determined by the costs of generating squeezed input states and the unitary entangling transformation. So, let's check the BS and CZ from this standpoint.

The energy cost associated with generating a squeezed state with the degree of squeezing $r$ \cite{MCTeich_1989} is typically estimated as
\BY
E = \sinh^{2}(r).
\label{n}
\EY
It is known that beam splitting, as well as the measurement of the number of particles, are not introduced energy costs. The energy costs $E_{BS}$ for generating the SF state in the scheme in Fig. \ref{fig:BS_CZ_gen}a are determined by the cost of the squeezing of two input oscillators:
\BY
E_{BS} = \sinh^{2}(r_{1})+\sinh^{2}(r_{2}).
\label{nbs}
\EY
Here, $r_{1}$ and $r_{2}$ are the squeezing degree of input oscillators in the first and second modes, respectively.

\begin{figure}[t]
    \centering
\includegraphics[width=1\linewidth]{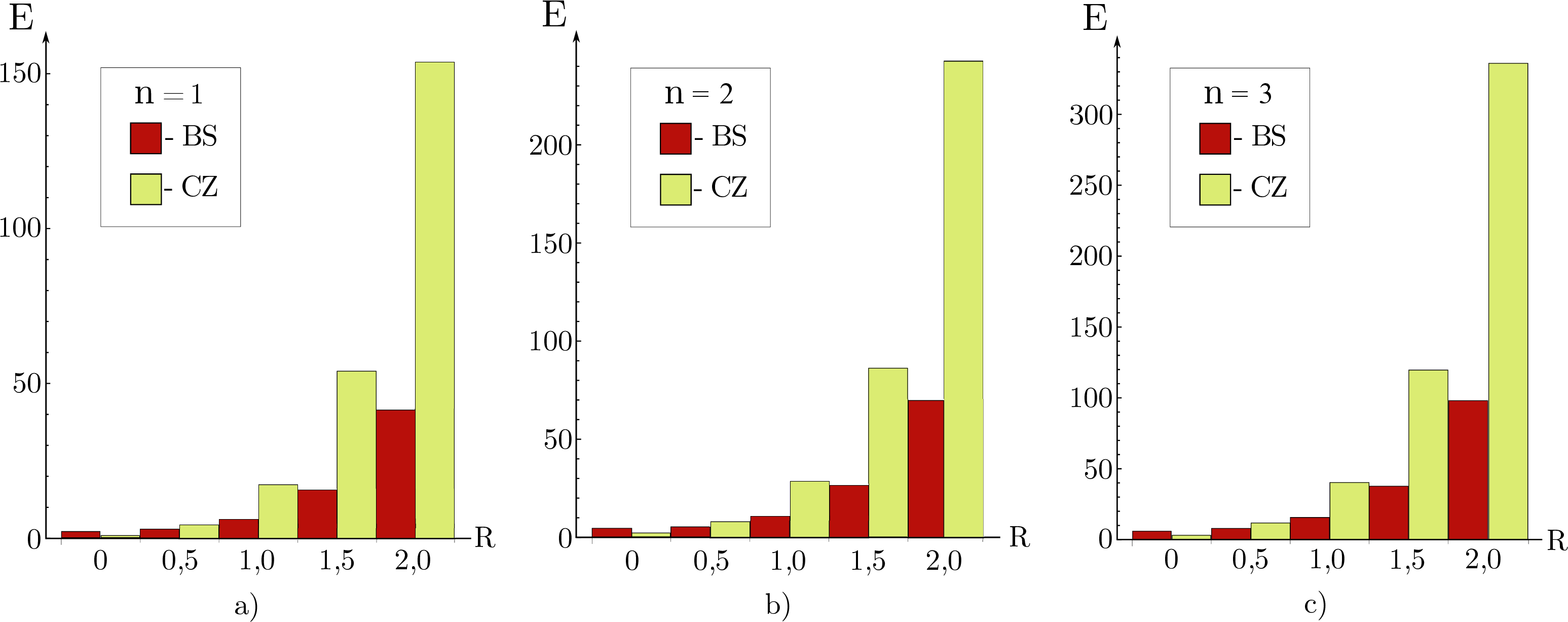}
    \caption{Histogram of the energy costs for generating squeezed Fock states with the degree of squeezing $R$ and a certain number $n$ (a, b, c at $n = 1, 2, 3$, respectively) in the protocol: with BS transformation (marked in dark burgundy color) and CZ gate (marked in light yellow color).}
    \label{ris:energylevel}
\end{figure}
\begin{figure}[t]
    \centering
    \includegraphics[scale=0.6]{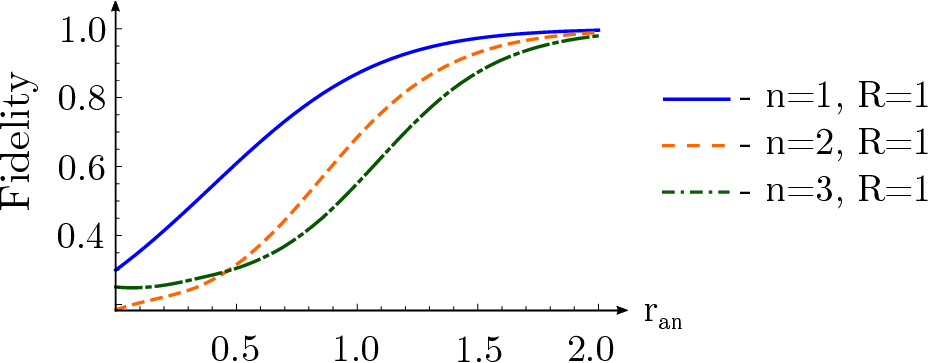}
    \caption{Dependence of the fidelity of generating the squeezed Fock states with certain numbers $n=1,2,3$ on the squeezing of auxiliary oscillators $r_{an}$ at a fixed squeezing degree of the output states $R=1$.}
\label{ris:fidelity}
\end{figure}

In Section II, we discussed the CZ gate in detail. The most significant characteristic of the CZ gate is the presence of two squeezed auxiliary oscillators. 
Strictly speaking, to achieve a perfect implementation of the CZ gate, the auxiliary oscillators must be perfectly squeezed. Accordingly, in the exact sense, this scheme requires infinite energy to squeeze the auxiliary oscillators. However, moving away from the concept of infinite squeezing to realistic systems, we can appreciate how sensitive the presented scheme is to the use of finitely squeezed oscillators as auxiliary ones. To achieve this, we calculated the fidelity of several realizable states with infinite and limited squeezing of auxiliary oscillators. Fig. \ref{ris:fidelity} shows the dependence of these functions on the ancillary squeezing degree.
The figure shows how auxiliary oscillators should be squeezed to generate a certain SF state with the required fidelity. By limiting the  fidelity to 0.99, we can estimate the energy costs for generation of the SF state.  Using the expressions (\ref{rg}) and (\ref{n}), we find that the energy costs are equal to
\BY
E_{CZ}(r_{1}, r_{2}, r_{an}) = \sinh^{2}(r_{1})&+\sinh^{2}(r_{2})+2 \sinh^{2}(r_{an}),
\label{ncz}
\EY
where the value of $r_{an}$ is chosen based on the requirement of generating a SF state with the fidelity of 0.99.
Using the BS and CZ setups, let us compare the energy costs required to generate SF states with a certain number $n$ and the degree of squeezing $R$. Let us emphasize again that we are comparing the exact generation in the BS case with the approximate generation in the CZ one. Fig. \ref{ris:energylevel} a, b, c demonstrates these results for generating SF states with numbers $n=1, 2, 3$, respectively. The $x$-axis represents the squeezing degree $R$ of SF states. The $y$-axis shows the energy costs required to generate the corresponding quantum states at two types of setups. In the Fig. \ref{ris:energylevel}, the dark burgundy color corresponds to the BS setup and the light yellow corresponds to the CZ case. As we can see, the energy costs for generating SF states increase exponentially with an increasing squeezing parameter $R$ for a given quantum state. It occurs in two types of setups. As expected, additional energy will be needed to generate SF states with a larger number. One can see that even the approximate generation of SF states in the CZ setup is less efficient than the exact generation in the BS case for all values of the target squeezing $R>0$. Since the BS setup losslessly converts the energy of the input states into the energy of the TMEG state before counting the photon number, this observation is not surprising. It should be noted that by reducing the requirement for fidelity to 0.95, it is possible to achieve conditions where, for $R<1$ , the use of the CZ setup becomes more energetically favorable.

In Section II, we discussed that to generate SF states with a certain number $n$ and  degree of squeezing $R$, it is necessary to satisfy the conditions of the universal solution regime. To generate certain SF states, the squeezing degree $r_{1}$ and $r_{2}$ of input states is not arbitrary, but it satisfies the conditions (\ref{cond0})-(\ref{cond1}). However, not any values of $r$ can be achieved experimentally. To date, the experimentally implemented squeezing limit is about $-15$ dB \cite{PhysRevLett.117.110801}. Therefore, we will analyze the maximum input squeezing degree required to generate SF states with certain parameters in the BS and CZ setups. It allows us to evaluate the possibility of different implementations of generating the SF states with a certain number $n$ and squeezing degree $R$ in a modern experiment.

\begin{figure}[H]
    \centering
    \includegraphics[width=1\linewidth]{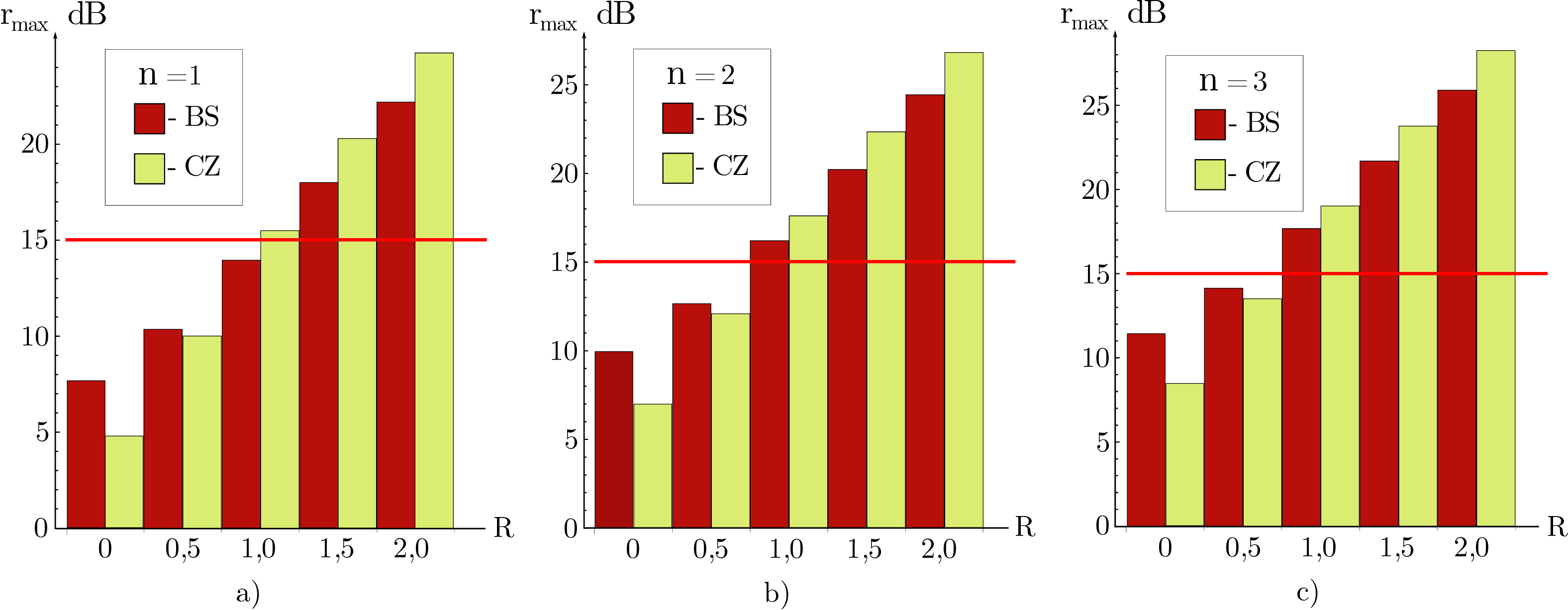}
    \caption{The maximum degree of input squeezing required to generate squeezed Fock states with the degree of squeezing $R$ and a certain number $n$ (a, b, c for $n=1, 2, 3$, respectively) in the protocol: with BS transformation (marked in dark burgundy color) and imperfect CZ gate with fidelity 0.99 (marked in light yellow color). In the figure, the red line indicates -15 dB that is the squeezing limit experimentally implemented to date.}
    \label{ris:sqlevel}
\end{figure}

To select the parameters that correspond to the universal solution regime, we conduct an analysis of the maximum required input squeezing to generate SF states with the squeezing parameter $R$ and the number $n$ (exact for the BS case and approximate with fidelity 0.99 for the CZ one). Furthermore, for the BS setup, there are two values, $r_{1}$ and $r_{2}$, from which we select the maximum. However, for the CZ gate, we have selected one of the maximum squeezing degree values out of the four required. Obtained results are presented in the Fig.\ref{ris:sqlevel}. The input squeezing degree is shown in dB. The red horizontal line indicates the experimentally achievable squeezing of $-15$ dB.  One can see that, for each setup, with an increasing squeezing parameter $R$, an increasingly higher input squeezing degree is required. One can be seen from the Fig. \ref{ris:sqlevel} that as the number $n$ of the SF state increases, the maximum required input squeezing degree also increases.  It is important to note that considering the CZ setup with the fidelity of 0.99, for $R<1$ (in light yellow color) requires less input squeezing (both among the input and auxiliary oscillators) compared to the BS setup (in dark burgundy color). It is clear from the obtained results that the CZ setup will not allow us to generate non-trivial $(n \geq 1)$ SF states with $R \geq 1$, whereas the BS configuration enables the generation of SF states with $R = 1$ and $n=1$.

Thus, the obtained results prove that, given the required squeezing resource to generate SF states, it is more profitable to use the BS setup where the required state is generated not only accurately, but also with minimal energy costs.

\section{Conclusion}
In this paper, we considered the protocol for generating the squeezed Fock (SF) states. This protocol can be implemented effectively using two entanglement operations: a beam splitter (BS) transformation and a CZ gate. The first transformation is passively interfered two beams. The second one contains a squeezing operation.  We compared these operations to generate the class of non-Gaussian states under consideration.

Generating a non-Gaussian state comprises several stages: generating initial squeezed states, performing a unitary entangling transformation, and measuring the number of particles in one of the output modes. Furthermore, any experimental implementation contains imperfections that affect the generated non-Gaussian states. Therefore, it was important to investigate how these aspects impact generating SF states.

We have checked the impact of losses on the generation of non-Gaussian states. Considering the limit of small uniform losses, we used the fact that it is possible to reduce the losses in the channels to the errors in the input squeezing.  For the BS setup, two input oscillators are subject to this effect. In the CZ case, the losses have an impact on both the input and auxiliary oscillators.  By linking losses in the system with the probabilistic process of reducing squeezing, we showed the resistance of the integral characteristics to losses. The BS setup has a smaller decrease in integral characteristics than the CZ case.

We examined the effect of the quantum efficiency of the detector on the accuracy of generating non-Gaussian states by measuring the number of particles in one mode. We have shown that the CZ setup has an advantage over the BS case in terms of the sensitivity of the fidelity of generating SF states to the quantum efficiency of PNRD. We have demonstrated that the fidelity of generating non-Gaussian states is independent of the squeezing degree of the generated state. The quantum efficiency $\eta_{eff}$ determines the accuracy of interpreting the number $n$. On the other hand, the accuracy in satisfaction of the requirements for the input parameters is responsible for the exactness of the squeezing parameter $R$ of the generated state. Thus, we managed to determine which parameters of the generated states were affected by different experimental imperfections. This is important since, for various applications, requirements for non-Gaussian states address to its different parameters. For example, the error correction protocols demand the correct squeezing of the non-Gaussian state, whereas quantum metrology desires the exact number of the Fock state.

Finally, we analyzed the energy cost of generating SF states using BS and imperfect (with fidelity equals 0.99) CZ setups. In the CZ case, energy costs include the costs of squeezing initial and auxiliary states. We have shown that these energy costs are always higher than the corresponding costs in the BS case. This is true for all target states with $R>0$. We have demonstrated that for $R>1$ the CZ setup has higher input state squeezing requirements than the BS case. When generating SF states with $R<1$, if the fidelity requirement is below 0.99, then the CZ setup has more lenient conditions on the required squeezing degree of the oscillators.

Our results showed that both schemes have their advantages and disadvantages. Depending on the experimental conditions, it may be preferable to use one or the other configuration. If the requirement for the fidelity of the target state exceeds 0.99, it is always better to use a  beam splitter setup.
 However, in this case, the requirement on the quality of the detector has significantly increased. Since the detection efficiency looks like a bottleneck of these schemes, both configurations are competitive.

\section*{Funding}
This research was supported by the Russian Science Foundation (Grant No. 24-22-00318).

\section*{Disclosures}
The authors declare no conflicts of interest.

\section*{Data availability} 
Data underlying the results presented in this paper are not publicly available at this time, but may be obtained from the authors upon reasonable request.

\appendix

\section{Fidelity between the generated state using the imperfect PNRD and the squeezed Fock state} \label{PNRD}  
Within this Appendix, we would like to give a technique for calculation of the fidelity (\ref{fidelity_ffin}) between the generated state at the quantum efficiency of the detector $\eta_{eff}\textless1$ and the squeezed Fock state. $\eta_{eff}\textless1$ is considered by introducing as an equivalent beam splitter, which entangles the measured mode with the vacuum one (see Fig. \ref{fig:PNDR}).

The general non-displaced TMEG state is described by the wave function shown in \cite{PhysRevA.109.052428}, as
\begin{align}
\label{TMGS}
    \Psi \left(x_1,x_2 \right)=\frac{\left(\mathrm{Re}[ a] \mathrm{Re} [d]-\left(\mathrm{Re }[b] \right)^2\right)^{1/4}}{\sqrt{\pi}} \exp \left[{-\frac{1}{2} \left(a x_1^2+2 b x_1 x_2+d x_2^2\right)}\right].
\end{align}

Considering all this, we can write the state of the system before the beam splitter (in Fig. \ref{fig:PNDR}) in the following form:
\begin{align}
    |\Psi\rangle  =\int \int\int  dx_1 dx_2 dx_3\Psi \left(x_1,x_2\right) \Psi_{vac} \left(x_3\right)|x_1\rangle _1 |x_2\rangle _2|x_3\rangle _3.
\end{align}
Here, $\Psi_{vac} \left(x_3\right)$ is the wave function of the vacuum state in the mode 3.

Next, the first  and vacuum third modes are mixed on the beam splitter with the energy transmittance $\eta_{eff}$. After the beam splitter transformation, the common 3-mode state is described as follows
\begin{align}
   |\Psi^{'}\rangle &=\int \int \int  dx_1 dx_2 dx_3 \Psi \left(\sqrt{\rho} \;x_{1}+\sqrt{\eta_{eff}}\;x_{3},x_{2}\right) \Psi_{vac} \left(\sqrt{\eta_{eff}}\;x_{1}-\sqrt{\rho}\;x_{3}\right)|x_1\rangle _1 |x_2\rangle _2|x_3\rangle _3.
\end{align}
Here, $\rho=1-\eta_{eff}$ is the energy reflection coefficient.

To study the dependence of the fidelity of the generated state on the quantum efficiency PNRD, it will be convenient for us to use the density matrix formalism. Then partially tracing out the vacuum  subsystem results in the reduced density matrix: 
\begin{align}  \label{f3}
&\hat{\rho}^{'}= Sp_{3}(\hat{\rho})=\int dx^{'}_{3}\;{}_{3}\langle x^{'}_{3} |\hat{\rho}|x^{'}_{3}\rangle{}_{3} =  \int dx^{'}_{3}\int\;dx_{1}\;dx_{2}\;d\tilde{x}_{1}\;d\tilde{x}_{2}\;\Psi \left(\sqrt{\rho} \;x_{1}+\sqrt{\eta_{eff}}\;x^{'}_{3},x_{2}\right)\;\Psi^{*} \left(\sqrt{\rho} \;\tilde{x}_{1}+\sqrt{\eta_{eff}}\;x^{'}_{3},\tilde{x}_{2}\right)\nonumber\\&\times\Psi_{vac} \left(\sqrt{\eta_{eff}}\;x_{1}-\sqrt{\rho}\;x^{'}_{3}\right)\Psi^{*}_{vac} \left(\sqrt{\eta_{eff}}\;\tilde{x}_{1}-\sqrt{\rho}\;x^{'}_{3}\right)\;|x_{1}\rangle{}_{1}|x_{2}\rangle{}_{2}\;{}_{1}\langle \tilde{x}_{1} |\;{}_{2}\langle \tilde{x}_{2}|.
\end{align}
Here, $\hat{\rho}$ is the density matrix operator of the three-mode state. The number of integrals is equal to the number of differentials.

Next, the state in the first mode is measured by the PNRD. Let us assume that the result of the measurement is the number $n$. Such a measurement leads to the following non-normalized state density matrix:
\begin{align} \label{rout}
&{}_{1}\langle n |\hat{\rho}^{'}|n\rangle{}_{1} 
=\int  dx^{'}_1\;dx^{''}_1 dx_1 dx_2 \;d\tilde{x}_{1}\;d\tilde{x}_{2} \;dx^{'}_3\;\Psi \left(\sqrt{\rho} \;x_{1}+\sqrt{\eta_{eff}}\;x^{'}_{3},x_{2}\right)\;\Psi^{*} \left(\sqrt{\rho} \;\tilde{x}_{1}+\sqrt{\eta_{eff}}\;x^{'}_{3},\tilde{x}_{2}\right)\nonumber\\&\times\Psi_{vac} \left(\sqrt{\eta_{eff}}\;x_{1}-\sqrt{\rho}\;x^{'}_{3}\right)\Psi^{*}_{vac} \left(\sqrt{\eta_{eff}}\;\tilde{x}_{1}-\sqrt{\rho}\;x^{'}_{3}\right)\;{}_1\langle n|x_1\rangle _1 |x_2\rangle _2\;{}_{1}\langle \tilde{x}_{1} |\;{}_{2}\langle \tilde{x}_{2}|n\rangle{}_{1} . 
\end{align}
Given that
$${}_1\langle n|x_1\rangle _1=\varphi_{n}(x_{1})=\frac{1}{\sqrt[^4]{\pi}2^{n/2}\sqrt{n!}}e^{-x_1^2/2}H_n(x_1) ,$$  
and $H_n(x)$ is the Hermite polynomial, we can write  the density matrix of the output state as:
\begin{align}  \label{int_out}
&\hat{\rho}_{out}= \frac{1}{\sqrt{N}}\int
  dx^{'}_1\;dx^{''}_1 \;dx_2\;\;d\tilde{x}_{2}\;dx^{'}_3 \;\Psi \left(\sqrt{\rho} \;x^{'}_{1}+\sqrt{\eta_{eff}}\;x^{'}_{3},x_{2}\right)\Psi_{vac} \left(\sqrt{\eta_{eff}}\;x^{'}_{1}-\sqrt{\rho}\;x^{'}_{3}\right)\varphi_{n}(x^{'}_{1})\nonumber\\ &\times \Psi^{*} \left(\sqrt{\rho} \;x^{''}_{1}+\sqrt{\eta_{eff}}\;x^{'}_{3},\tilde{x}_{2}\right)\Psi^{*}_{vac} \left(\sqrt{\eta_{eff}}\;x^{''}_{1}-\sqrt{\rho}\;x^{'}_{3}\right)\varphi^{*}_{n}(x^{''}_{1})=\int \;dx_2\;\;d\tilde{x}_{2}\;dx^{'}_3 \;\Psi_{out}(x_2, x^{'}_3,\eta_{eff})\;\nonumber\\ &\times\Psi^{*}_{out}(\tilde{x}_2, x^{'}_3,\eta_{eff}),
\end{align}
where
\begin{align}
\Psi_{out}(x_2, x^{'}_3,\eta_{eff})=\int dx^{'}_{1}\;\Psi \left(\sqrt{\rho} \;x^{'}_{1}+\sqrt{\eta_{eff}}\;x^{'}_{3},x_{2}\right)\Psi_{vac} \left(\sqrt{\eta_{eff}}\;x^{'}_{1}-\sqrt{\rho}\;x^{'}_{3}\right)\varphi_{n}(x^{'}_{1}),
\end{align}
and $N=\int dx_2 dx^{'}_3|\Psi_{out}(x_2, x^{;}_3,\eta_{eff})|^2 $ is the normalization of the output state density matrix.

Using the definition of fidelity (\ref{fidelity}), we can write:
\begin{align}  \label{f2}
  &F_{n}(\eta_{eff})=\int dx^{'}_{3}\;dx^{'}_{2}\;dx^{''}_{2}\;\Psi_{out}(x^{'}_2, x^{'}_3,\eta_{eff})\Psi_{\mathrm{\mathrm{SF}}}(x^{'}_{2},R, n)\Psi^{*}_{out}(x^{''}_{2}, x^{'}_{3},\eta_{eff})\Psi_{\mathrm{\mathrm{SF}}}^{*}(x^{''}_{2},R, n)=\nonumber\\&=\int dx^{'}_{3}\left|\int dx^{'}_{2}\Psi_{out}(x^{'}_2, x^{'}_3,\eta_{eff})\Psi_{\mathrm{\mathrm{SF}}}(x^{'}_{2}, R, n)\right|^{2}.
 \end{align}

\section{Quantum efficiency $\eta_{eff}\textless1$ of the detector in the protocol for generating non-Gaussian states by the measurement in the Fock representation}\label{POVM}

In this Appendix,  we would like to present a proof that the fidelity of the SF state generated in scheme Fig. \ref{fig:PNDR} does not depend on the squeezing parameters of the generated state at the quantum efficiency of the detector $\eta_{eff} \textless1$.

The imperfect PNRD in Fig. \ref{fig:PNDR} is taken into account by introducing as an equivalent beam splitter, which entangles the measured mode with the vacuum one. Assume $ |\Psi\rangle_{12} $ describes the entangled state of two modes (before the interaction with the loss channel). This state can be represented as an expansion on a Fock states basis:
\BY
|\Psi\rangle_{12}=\sum_{mn}C_{mn}|m\rangle_{1}|n\rangle_{2}.
\EY
Here, $C_{mn}$ are the expansion coefficients.

Let the quantum efficiency of the detector is   $\eta_{eff}=1$. Next, the state in the first channel is measured on the PNRD. Let us assume that the result of the measurement is the number  $M$. Such a measurement leads to the following output state vector:
\BY
&|\Psi_{M}^{\text{out}}\rangle_{2}={}_1\langle M|\sum_{mn}C_{mn}|m\rangle_{1}|n\rangle_{2}=\sum_{n}C_{Mn}|n\rangle_{2}.\quad\quad
\EY

But we want to consider that $\eta_{eff} \textless1$. To achieve this, we describe the entanglement of the first and vacuum third modes (see Fig. \ref{fig:PNDR}) in terms of the Fock states \cite{PhysRevA.55.3184,Clausen:1999kb}:
\BY
|m\rangle_{1}|0\rangle_{3}\xrightarrow{\hat{BS}_{13}}\sum_{k=0}^{m}(-1)^{k}&\sqrt{\frac{m!}{(m-k)!k!}}\eta_{eff}^{k}(1-\eta_{eff})^{m-k}|m-k\rangle_{1}|k\rangle_{3},\quad
\EY
where $\eta_{eff}$ and $1-\eta_{eff}$ are the energy transmission and reflection coefficients of the beam splitter, respectively. After the beam splitter transformation, the state of the system gets the following form:
\BY
&\hat{BS}_{13}|\Psi\rangle_{12}|0\rangle_{3}=\sum_{mn} C_{mn}\sum_{k=0}^{m}(-1)^{k}\sqrt{\frac{m!}{(m-k)!k!}}\eta_{eff}^{k}(1-\eta_{eff})^{m-k}|m-k\rangle_{1}|n\rangle_{2}|k\rangle_{3}.\quad
\EY

Next, the state in the first mode is measured on the detector. The measurement leads to the following state vector:
\BY \label{app_sup_b5}
&{}_1\langle M|\hat{BS}_{13}|\Psi\rangle_{12}|0\rangle_{3}=\sum_{mn}A_{mM}C_{mn}|n\rangle_{2}|m-M\rangle_{3}=\sum_{m}A_{mM}|m-M\rangle_{3}\sum_{n}C_{mn}|n\rangle_{2}\nonumber\\&=\sum_{m}A_{mM}|m-M\rangle_{3}|\Psi_{m}^{\text{out}}\rangle_{2}.
\label{R_final}
\EY
The decomposition coefficients $A_{mM}$ in Eq. (\ref{app_sup_b5}) have the following form:
\begin{align} \label{coef_AmM}
    A_{mM}=(-1)^{m-M}\sqrt{\frac{m!}{(m-M)!M!}}\eta_{eff}^{m-M}(1-\eta_{eff})^{M}.
\end{align}

These coefficients describe the probability that $M$ photons were detected by the PNRD, and $m$ is the number of photons subtracted from the state.  According to the obtained relationship, the quantum efficiency of the PNRD only affects  the decomposition coefficients $A_{mM}$. The state $|\Psi_{m}^{\text{out}}\rangle$ is independent of quantum efficiency.

According to \cite{korolev2024estimationsetstatesobtained}, the state $|\Psi_{m}^{\text{out}}\rangle$ represents by a finite superposition of the Fock states, which is acted upon by the squeezing operator. Only the measured number of photons affects the superposition, not the squeezing parameter. The scheme parameters totally determine the squeezing parameter. This means that the squeezing operator is independent of the summation index and can be factored out of the summation in (\ref{app_sup_b5}). Thus, the fidelity described by the expression (\ref{fidelity_ffin}) is independent of the squeezing parameter of the generated state.


\begin{thebibliography}{50}%
	\makeatletter
	\providecommand \@ifxundefined [1]{%
		\@ifx{#1\undefined}
	}%
	\providecommand \@ifnum [1]{%
		\ifnum #1\expandafter \@firstoftwo
		\else \expandafter \@secondoftwo
		\fi
	}%
	\providecommand \@ifx [1]{%
		\ifx #1\expandafter \@firstoftwo
		\else \expandafter \@secondoftwo
		\fi
	}%
	\providecommand \natexlab [1]{#1}%
	\providecommand \enquote  [1]{``#1''}%
	\providecommand \bibnamefont  [1]{#1}%
	\providecommand \bibfnamefont [1]{#1}%
	\providecommand \citenamefont [1]{#1}%
	\providecommand \href@noop [0]{\@secondoftwo}%
	\providecommand \href [0]{\begingroup \@sanitize@url \@href}%
	\providecommand \@href[1]{\@@startlink{#1}\@@href}%
	\providecommand \@@href[1]{\endgroup#1\@@endlink}%
	\providecommand \@sanitize@url [0]{\catcode `\\12\catcode `\$12\catcode
		`\&12\catcode `\#12\catcode `\^12\catcode `\_12\catcode `\%12\relax}%
	\providecommand \@@startlink[1]{}%
	\providecommand \@@endlink[0]{}%
	\providecommand \url  [0]{\begingroup\@sanitize@url \@url }%
	\providecommand \@url [1]{\endgroup\@href {#1}{\urlprefix }}%
	\providecommand \urlprefix  [0]{URL }%
	\providecommand \Eprint [0]{\href }%
	\providecommand \doibase [0]{https://doi.org/}%
	\providecommand \selectlanguage [0]{\@gobble}%
	\providecommand \bibinfo  [0]{\@secondoftwo}%
	\providecommand \bibfield  [0]{\@secondoftwo}%
	\providecommand \translation [1]{[#1]}%
	\providecommand \BibitemOpen [0]{}%
	\providecommand \bibitemStop [0]{}%
	\providecommand \bibitemNoStop [0]{.\EOS\space}%
	\providecommand \EOS [0]{\spacefactor3000\relax}%
	\providecommand \BibitemShut  [1]{\csname bibitem#1\endcsname}%
	\let\auto@bib@innerbib\@empty
	\bibitem [{\citenamefont {Braunstein}\ and\ \citenamefont {van
			Loock}(2005)}]{Braunstein_2005}%
	\BibitemOpen
	\bibfield  {author} {\bibinfo {author} {\bibfnamefont {S.~L.}\ \bibnamefont
			{Braunstein}}\ and\ \bibinfo {author} {\bibfnamefont {P.}~\bibnamefont {van
				Loock}},\ }\bibfield  {title} {\bibinfo {title} {Quantum information with
			continuous variables},\ }\href {https://doi.org/10.1103/RevModPhys.77.513}
	{\bibfield  {journal} {\bibinfo  {journal} {Rev. Mod. Phys.}\ }\textbf
		{\bibinfo {volume} {77}},\ \bibinfo {pages} {513} (\bibinfo {year}
		{2005})}\BibitemShut {NoStop}%
	\bibitem [{\citenamefont {Lloyd}\ and\ \citenamefont
		{Braunstein}(1999)}]{Lloyd_1999}%
	\BibitemOpen
	\bibfield  {author} {\bibinfo {author} {\bibfnamefont {S.}~\bibnamefont
			{Lloyd}}\ and\ \bibinfo {author} {\bibfnamefont {S.~L.}\ \bibnamefont
			{Braunstein}},\ }\bibfield  {title} {\bibinfo {title} {Quantum computation
			over continuous variables},\ }\href
	{https://doi.org/10.1103/PhysRevLett.82.1784} {\bibfield  {journal} {\bibinfo
			{journal} {Phys. Rev. Lett.}\ }\textbf {\bibinfo {volume} {82}},\ \bibinfo
		{pages} {1784} (\bibinfo {year} {1999})}\BibitemShut {NoStop}%
	\bibitem [{\citenamefont {Ralph}\ \emph {et~al.}(2003)\citenamefont {Ralph},
		\citenamefont {Gilchrist}, \citenamefont {Milburn}, \citenamefont {Munro},\
		and\ \citenamefont {Glancy}}]{Ralph_2003}%
	\BibitemOpen
	\bibfield  {author} {\bibinfo {author} {\bibfnamefont {T.~C.}\ \bibnamefont
			{Ralph}}, \bibinfo {author} {\bibfnamefont {A.}~\bibnamefont {Gilchrist}},
		\bibinfo {author} {\bibfnamefont {G.~J.}\ \bibnamefont {Milburn}}, \bibinfo
		{author} {\bibfnamefont {W.~J.}\ \bibnamefont {Munro}},\ and\ \bibinfo
		{author} {\bibfnamefont {S.}~\bibnamefont {Glancy}},\ }\bibfield  {title}
	{\bibinfo {title} {Quantum computation with optical coherent states},\ }\href
	{https://doi.org/10.1103/PhysRevA.68.042319} {\bibfield  {journal} {\bibinfo
			{journal} {Phys. Rev. A}\ }\textbf {\bibinfo {volume} {68}},\ \bibinfo
		{pages} {042319} (\bibinfo {year} {2003})}\BibitemShut {NoStop}%
	\bibitem [{\citenamefont {Hastrup}\ and\ \citenamefont
		{Andersen}(2022)}]{Hastrup_2022}%
	\BibitemOpen
	\bibfield  {author} {\bibinfo {author} {\bibfnamefont {J.}~\bibnamefont
			{Hastrup}}\ and\ \bibinfo {author} {\bibfnamefont {U.~L.}\ \bibnamefont
			{Andersen}},\ }\bibfield  {title} {\bibinfo {title} {All-optical cat-code
			quantum error correction},\ }\href
	{https://doi.org/10.1103/PhysRevResearch.4.043065} {\bibfield  {journal}
		{\bibinfo  {journal} {Phys. Rev. Res.}\ }\textbf {\bibinfo {volume} {4}},\
		\bibinfo {pages} {043065} (\bibinfo {year} {2022})}\BibitemShut {NoStop}%
	\bibitem [{\citenamefont {Zhang}\ \emph {et~al.}(2019)\citenamefont {Zhang},
		\citenamefont {Zhang}, \citenamefont {Hu}, \citenamefont {Cen}, \citenamefont
		{Sun}, \citenamefont {Jin},\ and\ \citenamefont {Zhao}}]{Zhang}%
	\BibitemOpen
	\bibfield  {author} {\bibinfo {author} {\bibfnamefont {J.-D.}\ \bibnamefont
			{Zhang}}, \bibinfo {author} {\bibfnamefont {Z.-J.}\ \bibnamefont {Zhang}},
		\bibinfo {author} {\bibfnamefont {J.-Y.}\ \bibnamefont {Hu}}, \bibinfo
		{author} {\bibfnamefont {L.-Z.}\ \bibnamefont {Cen}}, \bibinfo {author}
		{\bibfnamefont {Y.-F.}\ \bibnamefont {Sun}}, \bibinfo {author} {\bibfnamefont
			{C.-F.}\ \bibnamefont {Jin}},\ and\ \bibinfo {author} {\bibfnamefont
			{Y.}~\bibnamefont {Zhao}},\ }\bibfield  {title} {\bibinfo {title} {Conclusive
			nonlinear phase sensitivity limit for a mach-zehnder interferometer with
			single-mode non-vacuum inputs},\ }\href {https://arxiv.org/abs/1906.10867}
	{\bibfield  {journal} {\bibinfo  {journal} {arXiv}\ }\textbf {\bibinfo
			{volume} {1906.10867}} (\bibinfo {year} {2019})}\BibitemShut {NoStop}%
	\bibitem [{\citenamefont {Hou}\ \emph {et~al.}(2019)\citenamefont {Hou},
		\citenamefont {Sui}, \citenamefont {Wang},\ and\ \citenamefont {Xu}}]{Hou}%
	\BibitemOpen
	\bibfield  {author} {\bibinfo {author} {\bibfnamefont {L.-L.}\ \bibnamefont
			{Hou}}, \bibinfo {author} {\bibfnamefont {Y.-X.}\ \bibnamefont {Sui}},
		\bibinfo {author} {\bibfnamefont {S.}~\bibnamefont {Wang}},\ and\ \bibinfo
		{author} {\bibfnamefont {X.-F.}\ \bibnamefont {Xu}},\ }\bibfield  {title}
	{\bibinfo {title} {Quantum interferometry via a coherent state mixed with a
			squeezed number state*},\ }\href
	{https://doi.org/10.1088/1674-1056/28/4/044203} {\bibfield  {journal}
		{\bibinfo  {journal} {Chinese Physics B}\ }\textbf {\bibinfo {volume} {28}},\
		\bibinfo {pages} {044203} (\bibinfo {year} {2019})}\BibitemShut {NoStop}%
	\bibitem [{\citenamefont {S\'anchez Mu\~noz}\ \emph {et~al.}(2021)\citenamefont
		{S\'anchez Mu\~noz}, \citenamefont {Frascella},\ and\ \citenamefont
		{Schlawin}}]{PhysRevResearch.3.033250}%
	\BibitemOpen
	\bibfield  {author} {\bibinfo {author} {\bibfnamefont {C.}~\bibnamefont
			{S\'anchez Mu\~noz}}, \bibinfo {author} {\bibfnamefont {G.}~\bibnamefont
			{Frascella}},\ and\ \bibinfo {author} {\bibfnamefont {F.}~\bibnamefont
			{Schlawin}},\ }\bibfield  {title} {\bibinfo {title} {Quantum metrology of
			two-photon absorption},\ }\href
	{https://doi.org/10.1103/PhysRevResearch.3.033250} {\bibfield  {journal}
		{\bibinfo  {journal} {Phys. Rev. Res.}\ }\textbf {\bibinfo {volume} {3}},\
		\bibinfo {pages} {033250} (\bibinfo {year} {2021})}\BibitemShut {NoStop}%
	\bibitem [{\citenamefont {Lee}\ \emph {et~al.}(2019)\citenamefont {Lee},
		\citenamefont {Park},\ and\ \citenamefont {Nha}}]{Lee2019}%
	\BibitemOpen
	\bibfield  {author} {\bibinfo {author} {\bibfnamefont {J.}~\bibnamefont
			{Lee}}, \bibinfo {author} {\bibfnamefont {J.}~\bibnamefont {Park}},\ and\
		\bibinfo {author} {\bibfnamefont {H.}~\bibnamefont {Nha}},\ }\bibfield
	{title} {\bibinfo {title} {Quantum non-gaussianity and secure quantum
			communication},\ }\href {https://doi.org/10.1038/s41534-019-0164-9}
	{\bibfield  {journal} {\bibinfo  {journal} {npj Quantum Information}\
		}\textbf {\bibinfo {volume} {5}},\ \bibinfo {pages} {49} (\bibinfo {year}
		{2019})}\BibitemShut {NoStop}%
	\bibitem [{\citenamefont {Guo}\ \emph {et~al.}(2019)\citenamefont {Guo},
		\citenamefont {Ye}, \citenamefont {Zhong},\ and\ \citenamefont
		{Liao}}]{Guo2019}%
	\BibitemOpen
	\bibfield  {author} {\bibinfo {author} {\bibfnamefont {Y.}~\bibnamefont
			{Guo}}, \bibinfo {author} {\bibfnamefont {W.}~\bibnamefont {Ye}}, \bibinfo
		{author} {\bibfnamefont {H.}~\bibnamefont {Zhong}},\ and\ \bibinfo {author}
		{\bibfnamefont {Q.}~\bibnamefont {Liao}},\ }\bibfield  {title} {\bibinfo
		{title} {Continuous-variable quantum key distribution with non-gaussian
			quantum catalysis},\ }\href {https://doi.org/10.1103/PhysRevA.99.032327}
	{\bibfield  {journal} {\bibinfo  {journal} {Phys. Rev. A}\ }\textbf {\bibinfo
			{volume} {99}},\ \bibinfo {pages} {032327} (\bibinfo {year}
		{2019})}\BibitemShut {NoStop}%
	\bibitem [{\citenamefont {Opatrn\'y}\ \emph {et~al.}(2000)\citenamefont
		{Opatrn\'y}, \citenamefont {Kurizki},\ and\ \citenamefont
		{Welsch}}]{PhysRevA.61.032302}%
	\BibitemOpen
	\bibfield  {author} {\bibinfo {author} {\bibfnamefont {T.}~\bibnamefont
			{Opatrn\'y}}, \bibinfo {author} {\bibfnamefont {G.}~\bibnamefont {Kurizki}},\
		and\ \bibinfo {author} {\bibfnamefont {D.-G.}\ \bibnamefont {Welsch}},\
	}\bibfield  {title} {\bibinfo {title} {Improvement on teleportation of
			continuous variables by photon subtraction via conditional measurement},\
	}\href {https://doi.org/10.1103/PhysRevA.61.032302} {\bibfield  {journal}
		{\bibinfo  {journal} {Phys. Rev. A}\ }\textbf {\bibinfo {volume} {61}},\
		\bibinfo {pages} {032302} (\bibinfo {year} {2000})}\BibitemShut {NoStop}%
	\bibitem [{\citenamefont {Zinatullin}\ \emph {et~al.}(2023)\citenamefont
		{Zinatullin}, \citenamefont {Korolev},\ and\ \citenamefont
		{Golubeva}}]{Zinatullin2023}%
	\BibitemOpen
	\bibfield  {author} {\bibinfo {author} {\bibfnamefont {E.~R.}\ \bibnamefont
			{Zinatullin}}, \bibinfo {author} {\bibfnamefont {S.~B.}\ \bibnamefont
			{Korolev}},\ and\ \bibinfo {author} {\bibfnamefont {T.~Y.}\ \bibnamefont
			{Golubeva}},\ }\bibfield  {title} {\bibinfo {title} {Teleportation protocols
			with non-gaussian operations: Conditional photon subtraction versus cubic
			phase gate},\ }\href {https://doi.org/10.1103/PhysRevA.107.022422} {\bibfield
		{journal} {\bibinfo  {journal} {Phys. Rev. A}\ }\textbf {\bibinfo {volume}
			{107}},\ \bibinfo {pages} {022422} (\bibinfo {year} {2023})}\BibitemShut
	{NoStop}%
	\bibitem [{\citenamefont {Zinatullin}\ \emph {et~al.}(2021)\citenamefont
		{Zinatullin}, \citenamefont {Korolev},\ and\ \citenamefont
		{Golubeva}}]{Zinatullin2021}%
	\BibitemOpen
	\bibfield  {author} {\bibinfo {author} {\bibfnamefont {E.~R.}\ \bibnamefont
			{Zinatullin}}, \bibinfo {author} {\bibfnamefont {S.~B.}\ \bibnamefont
			{Korolev}},\ and\ \bibinfo {author} {\bibfnamefont {T.~Y.}\ \bibnamefont
			{Golubeva}},\ }\bibfield  {title} {\bibinfo {title} {Teleportation with a
			cubic phase gate},\ }\href {https://doi.org/10.1103/PhysRevA.104.032420}
	{\bibfield  {journal} {\bibinfo  {journal} {Phys. Rev. A}\ }\textbf {\bibinfo
			{volume} {104}},\ \bibinfo {pages} {032420} (\bibinfo {year}
		{2021})}\BibitemShut {NoStop}%
	\bibitem [{\citenamefont {Kienzler}\ \emph {et~al.}(2017)\citenamefont
		{Kienzler}, \citenamefont {Lo}, \citenamefont {Negnevitsky}, \citenamefont
		{Fl\"uhmann}, \citenamefont {Marinelli},\ and\ \citenamefont
		{Home}}]{PhysRevLett.119.033602}%
	\BibitemOpen
	\bibfield  {author} {\bibinfo {author} {\bibfnamefont {D.}~\bibnamefont
			{Kienzler}}, \bibinfo {author} {\bibfnamefont {H.-Y.}\ \bibnamefont {Lo}},
		\bibinfo {author} {\bibfnamefont {V.}~\bibnamefont {Negnevitsky}}, \bibinfo
		{author} {\bibfnamefont {C.}~\bibnamefont {Fl\"uhmann}}, \bibinfo {author}
		{\bibfnamefont {M.}~\bibnamefont {Marinelli}},\ and\ \bibinfo {author}
		{\bibfnamefont {J.~P.}\ \bibnamefont {Home}},\ }\bibfield  {title} {\bibinfo
		{title} {Quantum harmonic oscillator state control in a squeezed fock
			basis},\ }\href {https://doi.org/10.1103/PhysRevLett.119.033602} {\bibfield
		{journal} {\bibinfo  {journal} {Phys. Rev. Lett.}\ }\textbf {\bibinfo
			{volume} {119}},\ \bibinfo {pages} {033602} (\bibinfo {year}
		{2017})}\BibitemShut {NoStop}%
	\bibitem [{\citenamefont {Fiur\'a\ifmmode~\check{s}\else \v{s}\fi{}ek}\ and\
		\citenamefont {Je\ifmmode~\check{z}\else
			\v{z}\fi{}ek}(2013)}]{PhysRevA.87.062115}%
	\BibitemOpen
	\bibfield  {author} {\bibinfo {author} {\bibfnamefont {J.}~\bibnamefont
			{Fiur\'a\ifmmode~\check{s}\else s\fi{}ek}}\ and\ \bibinfo {author}
		{\bibfnamefont {M.}~\bibnamefont {Je\ifmmode~\check{z}\else z\fi{}ek}},\
	}\bibfield  {title} {\bibinfo {title} {Witnessing negativity of wigner
			function by estimating fidelities of catlike states from homodyne
			measurements},\ }\href {https://doi.org/10.1103/PhysRevA.87.062115}
	{\bibfield  {journal} {\bibinfo  {journal} {Phys. Rev. A}\ }\textbf {\bibinfo
			{volume} {87}},\ \bibinfo {pages} {062115} (\bibinfo {year}
		{2013})}\BibitemShut {NoStop}%
	\bibitem [{\citenamefont {Olivares}\ and\ \citenamefont
		{Paris}(2006)}]{Olivares2006}%
	\BibitemOpen
	\bibfield  {author} {\bibinfo {author} {\bibfnamefont {S.}~\bibnamefont
			{Olivares}}\ and\ \bibinfo {author} {\bibfnamefont {M.~G.~A.}\ \bibnamefont
			{Paris}},\ }\bibfield  {title} {\bibinfo {title} {De-gaussification by
			inconclusive photon subtraction},\ }\href
	{https://doi.org/10.1134/S1054660X06110077} {\bibfield  {journal} {\bibinfo
			{journal} {Laser Physics}\ }\textbf {\bibinfo {volume} {16}},\ \bibinfo
		{pages} {1533} (\bibinfo {year} {2006})}\BibitemShut {NoStop}%
	\bibitem [{\citenamefont {Olivares}\ and\ \citenamefont
		{Paris}(2005)}]{olivares2005squeezed}%
	\BibitemOpen
	\bibfield  {author} {\bibinfo {author} {\bibfnamefont {S.}~\bibnamefont
			{Olivares}}\ and\ \bibinfo {author} {\bibfnamefont {M.~G.}\ \bibnamefont
			{Paris}},\ }\bibfield  {title} {\bibinfo {title} {Squeezed fock state by
			inconclusive photon subtraction},\ }\href@noop {} {\bibfield  {journal}
		{\bibinfo  {journal} {Journal of Optics B: Quantum and Semiclassical Optics}\
		}\textbf {\bibinfo {volume} {7}},\ \bibinfo {pages} {S616} (\bibinfo {year}
		{2005})}\BibitemShut {NoStop}%
	\bibitem [{\citenamefont {Král}(1990)}]{Kral1990}%
	\BibitemOpen
	\bibfield  {author} {\bibinfo {author} {\bibfnamefont {P.}~\bibnamefont
			{Král}},\ }\bibfield  {title} {\bibinfo {title} {Displaced and squeezed fock
			states},\ }\href {https://doi.org/10.1080/09500349014550941} {\bibfield
		{journal} {\bibinfo  {journal} {Journal of Modern Optics}\ }\textbf {\bibinfo
			{volume} {37}},\ \bibinfo {pages} {889} (\bibinfo {year} {1990})},\ \Eprint
	{https://arxiv.org/abs/https://doi.org/10.1080/09500349014550941}
	{https://doi.org/10.1080/09500349014550941} \BibitemShut {NoStop}%
	\bibitem [{\citenamefont {Nieto}(1997)}]{NIETO1997135}%
	\BibitemOpen
	\bibfield  {author} {\bibinfo {author} {\bibfnamefont {M.~M.}\ \bibnamefont
			{Nieto}},\ }\bibfield  {title} {\bibinfo {title} {Displaced and squeezed
			number states},\ }\href
	{https://doi.org/https://doi.org/10.1016/S0375-9601(97)00183-7} {\bibfield
		{journal} {\bibinfo  {journal} {Physics Letters A}\ }\textbf {\bibinfo
			{volume} {229}},\ \bibinfo {pages} {135} (\bibinfo {year}
		{1997})}\BibitemShut {NoStop}%
	\bibitem [{\citenamefont {Kim}\ \emph {et~al.}(1989)\citenamefont {Kim},
		\citenamefont {de~Oliveira},\ and\ \citenamefont
		{Knight}}]{PhysRevA.40.2494}%
	\BibitemOpen
	\bibfield  {author} {\bibinfo {author} {\bibfnamefont {M.~S.}\ \bibnamefont
			{Kim}}, \bibinfo {author} {\bibfnamefont {F.~A.~M.}\ \bibnamefont
			{de~Oliveira}},\ and\ \bibinfo {author} {\bibfnamefont {P.~L.}\ \bibnamefont
			{Knight}},\ }\bibfield  {title} {\bibinfo {title} {Properties of squeezed
			number states and squeezed thermal states},\ }\href
	{https://doi.org/10.1103/PhysRevA.40.2494} {\bibfield  {journal} {\bibinfo
			{journal} {Phys. Rev. A}\ }\textbf {\bibinfo {volume} {40}},\ \bibinfo
		{pages} {2494} (\bibinfo {year} {1989})}\BibitemShut {NoStop}%
	\bibitem [{\citenamefont {Fiur\'a\ifmmode~\check{s}\else \v{s}\fi{}ek}\ \emph
		{et~al.}(2005)\citenamefont {Fiur\'a\ifmmode~\check{s}\else \v{s}\fi{}ek},
		\citenamefont {Garc\'{\i}a-Patr\'on},\ and\ \citenamefont
		{Cerf}}]{PhysRevA.72.033822}%
	\BibitemOpen
	\bibfield  {author} {\bibinfo {author} {\bibfnamefont {J.}~\bibnamefont
			{Fiur\'a\ifmmode~\check{s}\else s\fi{}ek}}, \bibinfo {author}
		{\bibfnamefont {R.}~\bibnamefont {Garc\'{\i}a-Patr\'on}},\ and\ \bibinfo
		{author} {\bibfnamefont {N.~J.}\ \bibnamefont {Cerf}},\ }\bibfield  {title}
	{\bibinfo {title} {Conditional generation of arbitrary single-mode quantum
			states of light by repeated photon subtractions},\ }\href
	{https://doi.org/10.1103/PhysRevA.72.033822} {\bibfield  {journal} {\bibinfo
			{journal} {Phys. Rev. A}\ }\textbf {\bibinfo {volume} {72}},\ \bibinfo
		{pages} {033822} (\bibinfo {year} {2005})}\BibitemShut {NoStop}%
	\bibitem [{\citenamefont {Neergaard-Nielsen}\ \emph {et~al.}(2010)\citenamefont
		{Neergaard-Nielsen}, \citenamefont {Takeuchi}, \citenamefont {Wakui},
		\citenamefont {Takahashi}, \citenamefont {Hayasaka}, \citenamefont
		{Takeoka},\ and\ \citenamefont {Sasaki}}]{PhysRevLett.105.053602}%
	\BibitemOpen
	\bibfield  {author} {\bibinfo {author} {\bibfnamefont {J.~S.}\ \bibnamefont
			{Neergaard-Nielsen}}, \bibinfo {author} {\bibfnamefont {M.}~\bibnamefont
			{Takeuchi}}, \bibinfo {author} {\bibfnamefont {K.}~\bibnamefont {Wakui}},
		\bibinfo {author} {\bibfnamefont {H.}~\bibnamefont {Takahashi}}, \bibinfo
		{author} {\bibfnamefont {K.}~\bibnamefont {Hayasaka}}, \bibinfo {author}
		{\bibfnamefont {M.}~\bibnamefont {Takeoka}},\ and\ \bibinfo {author}
		{\bibfnamefont {M.}~\bibnamefont {Sasaki}},\ }\bibfield  {title} {\bibinfo
		{title} {Optical continuous-variable qubit},\ }\href
	{https://doi.org/10.1103/PhysRevLett.105.053602} {\bibfield  {journal}
		{\bibinfo  {journal} {Phys. Rev. Lett.}\ }\textbf {\bibinfo {volume} {105}},\
		\bibinfo {pages} {053602} (\bibinfo {year} {2010})}\BibitemShut {NoStop}%
	\bibitem [{\citenamefont {Korolev}\ \emph {et~al.}(2023)\citenamefont
		{Korolev}, \citenamefont {Bashmakova},\ and\ \citenamefont
		{Golubeva}}]{QECCSFOCK}%
	\BibitemOpen
	\bibfield  {author} {\bibinfo {author} {\bibfnamefont {S.~B.}\ \bibnamefont
			{Korolev}}, \bibinfo {author} {\bibfnamefont {E.~N.}\ \bibnamefont
			{Bashmakova}},\ and\ \bibinfo {author} {\bibfnamefont {T.~Y.}\ \bibnamefont
			{Golubeva}},\ }\bibfield  {title} {\bibinfo {title} {Error correction using
			squeezed fock states},\ }\href {https://arxiv.org/abs/2312.16000} {\bibfield
		{journal} {\bibinfo  {journal} {arXiv}\ }\textbf {\bibinfo {volume}
			{2312.16000}} (\bibinfo {year} {2023})}\BibitemShut {NoStop}%
	\bibitem [{\citenamefont {Sychev}\ \emph {et~al.}(2017)\citenamefont {Sychev},
		\citenamefont {Ulanov}, \citenamefont {Pushkina}, \citenamefont {Richards},
		\citenamefont {Fedorov},\ and\ \citenamefont {Lvovsky}}]{Sychev2017}%
	\BibitemOpen
	\bibfield  {author} {\bibinfo {author} {\bibfnamefont {D.~V.}\ \bibnamefont
			{Sychev}}, \bibinfo {author} {\bibfnamefont {A.~E.}\ \bibnamefont {Ulanov}},
		\bibinfo {author} {\bibfnamefont {A.~A.}\ \bibnamefont {Pushkina}}, \bibinfo
		{author} {\bibfnamefont {M.~W.}\ \bibnamefont {Richards}}, \bibinfo {author}
		{\bibfnamefont {I.~A.}\ \bibnamefont {Fedorov}},\ and\ \bibinfo {author}
		{\bibfnamefont {A.~I.}\ \bibnamefont {Lvovsky}},\ }\bibfield  {title}
	{\bibinfo {title} {Enlargement of optical schr{\"o}dinger's cat states},\
	}\href {https://doi.org/10.1038/nphoton.2017.57} {\bibfield  {journal}
		{\bibinfo  {journal} {Nature Photonics}\ }\textbf {\bibinfo {volume} {11}},\
		\bibinfo {pages} {379} (\bibinfo {year} {2017})}\BibitemShut {NoStop}%
	\bibitem [{\citenamefont {Bužek}\ and\ \citenamefont
		{Knight}(1995)}]{Buzek1995}%
	\BibitemOpen
	\bibfield  {author} {\bibinfo {author} {\bibfnamefont {V.}~\bibnamefont
			{Buzek}}\ and\ \bibinfo {author} {\bibfnamefont {P.~L.}\ \bibnamefont
			{Knight}},\ }\bibfield  {title} {\bibinfo {title} {I: Quantum interference,
			superposition states of light, and nonclassical effects}\ }(\bibinfo
	{publisher} {Elsevier},\ \bibinfo {year} {1995})\ pp.\ \bibinfo {pages}
	{1--158}\BibitemShut {NoStop}%
	\bibitem [{\citenamefont {Bashmakova}\ \emph {et~al.}(2023)\citenamefont
		{Bashmakova}, \citenamefont {Korolev},\ and\ \citenamefont
		{Golubeva}}]{bashmakova2023effect}%
	\BibitemOpen
	\bibfield  {author} {\bibinfo {author} {\bibfnamefont {E.}~\bibnamefont
			{Bashmakova}}, \bibinfo {author} {\bibfnamefont {S.}~\bibnamefont
			{Korolev}},\ and\ \bibinfo {author} {\bibfnamefont {T.~Y.}\ \bibnamefont
			{Golubeva}},\ }\bibfield  {title} {\bibinfo {title} {Effect of entanglement
			in the generalized photon subtraction scheme},\ }\href@noop {} {\bibfield
		{journal} {\bibinfo  {journal} {Laser Physics Letters}\ }\textbf {\bibinfo
			{volume} {20}},\ \bibinfo {pages} {115203} (\bibinfo {year}
		{2023})}\BibitemShut {NoStop}%
	\bibitem [{\citenamefont {Ourjoumtsev}\ \emph {et~al.}(2007)\citenamefont
		{Ourjoumtsev}, \citenamefont {Dantan}, \citenamefont {Tualle-Brouri},\ and\
		\citenamefont {Grangier}}]{Ourjoumtsev2007}%
	\BibitemOpen
	\bibfield  {author} {\bibinfo {author} {\bibfnamefont {A.}~\bibnamefont
			{Ourjoumtsev}}, \bibinfo {author} {\bibfnamefont {A.}~\bibnamefont {Dantan}},
		\bibinfo {author} {\bibfnamefont {R.}~\bibnamefont {Tualle-Brouri}},\ and\
		\bibinfo {author} {\bibfnamefont {P.}~\bibnamefont {Grangier}},\ }\bibfield
	{title} {\bibinfo {title} {Increasing entanglement between gaussian states by
			coherent photon subtraction},\ }\href
	{https://doi.org/10.1103/PhysRevLett.98.030502} {\bibfield  {journal}
		{\bibinfo  {journal} {Phys. Rev. Lett.}\ }\textbf {\bibinfo {volume} {98}},\
		\bibinfo {pages} {030502} (\bibinfo {year} {2007})}\BibitemShut {NoStop}%
	\bibitem [{\citenamefont {Dell'Anno}\ \emph {et~al.}(2018)\citenamefont
		{Dell'Anno}, \citenamefont {Buono}, \citenamefont {Nocerino}, \citenamefont
		{De~Siena},\ and\ \citenamefont {Illuminati}}]{DellAnno}%
	\BibitemOpen
	\bibfield  {author} {\bibinfo {author} {\bibfnamefont {F.}~\bibnamefont
			{Dell'Anno}}, \bibinfo {author} {\bibfnamefont {D.}~\bibnamefont {Buono}},
		\bibinfo {author} {\bibfnamefont {G.}~\bibnamefont {Nocerino}}, \bibinfo
		{author} {\bibfnamefont {S.}~\bibnamefont {De~Siena}},\ and\ \bibinfo
		{author} {\bibfnamefont {F.}~\bibnamefont {Illuminati}},\ }\bibfield  {title}
	{\bibinfo {title} {Non-gaussian swapping of entangled resources},\ }\href
	{https://doi.org/10.1007/s11128-018-2133-1} {\bibfield  {journal} {\bibinfo
			{journal} {Quantum Information Processing}\ }\textbf {\bibinfo {volume}
			{18}},\ \bibinfo {pages} {20} (\bibinfo {year} {2018})}\BibitemShut {NoStop}%
	\bibitem [{\citenamefont {Winnel}\ \emph {et~al.}(2024)\citenamefont {Winnel},
		\citenamefont {Guanzon}, \citenamefont {Singh},\ and\ \citenamefont
		{Ralph}}]{PhysRevLett.132.230602}%
	\BibitemOpen
	\bibfield  {author} {\bibinfo {author} {\bibfnamefont {M.~S.}\ \bibnamefont
			{Winnel}}, \bibinfo {author} {\bibfnamefont {J.~J.}\ \bibnamefont {Guanzon}},
		\bibinfo {author} {\bibfnamefont {D.}~\bibnamefont {Singh}},\ and\ \bibinfo
		{author} {\bibfnamefont {T.~C.}\ \bibnamefont {Ralph}},\ }\bibfield  {title}
	{\bibinfo {title} {Deterministic preparation of optical squeezed cat and
			gottesman-kitaev-preskill states},\ }\href
	{https://doi.org/10.1103/PhysRevLett.132.230602} {\bibfield  {journal}
		{\bibinfo  {journal} {Phys. Rev. Lett.}\ }\textbf {\bibinfo {volume} {132}},\
		\bibinfo {pages} {230602} (\bibinfo {year} {2024})}\BibitemShut {NoStop}%
	\bibitem [{\citenamefont {Podoshvedov}\ \emph {et~al.}(2023)\citenamefont
		{Podoshvedov}, \citenamefont {Podoshvedov},\ and\ \citenamefont
		{Kulik}}]{Podoshvedov_2023}%
	\BibitemOpen
	\bibfield  {author} {\bibinfo {author} {\bibfnamefont {M.~S.}\ \bibnamefont
			{Podoshvedov}}, \bibinfo {author} {\bibfnamefont {S.~A.}\ \bibnamefont
			{Podoshvedov}},\ and\ \bibinfo {author} {\bibfnamefont {S.~P.}\ \bibnamefont
			{Kulik}},\ }\bibfield  {title} {\bibinfo {title} {Algorithm of quantum
			engineering of large-amplitude high-fidelity schr{\"o}dinger cat states},\
	}\href {https://doi.org/10.1038/s41598-023-30218-6} {\bibfield  {journal}
		{\bibinfo  {journal} {Scientific Reports}\ }\textbf {\bibinfo {volume}
			{13}},\ \bibinfo {pages} {3965} (\bibinfo {year} {2023})}\BibitemShut
	{NoStop}%
	\bibitem [{\citenamefont {Takase}\ \emph {et~al.}(2021)\citenamefont {Takase},
		\citenamefont {Yoshikawa}, \citenamefont {Asavanant}, \citenamefont {Endo},\
		and\ \citenamefont {Furusawa}}]{Takase2021}%
	\BibitemOpen
	\bibfield  {author} {\bibinfo {author} {\bibfnamefont {K.}~\bibnamefont
			{Takase}}, \bibinfo {author} {\bibfnamefont {J.-i.}\ \bibnamefont
			{Yoshikawa}}, \bibinfo {author} {\bibfnamefont {W.}~\bibnamefont
			{Asavanant}}, \bibinfo {author} {\bibfnamefont {M.}~\bibnamefont {Endo}},\
		and\ \bibinfo {author} {\bibfnamefont {A.}~\bibnamefont {Furusawa}},\
	}\bibfield  {title} {\bibinfo {title} {Generation of optical schr\"odinger
			cat states by generalized photon subtraction},\ }\href
	{https://doi.org/10.1103/PhysRevA.103.013710} {\bibfield  {journal} {\bibinfo
			{journal} {Phys. Rev. A}\ }\textbf {\bibinfo {volume} {103}},\ \bibinfo
		{pages} {013710} (\bibinfo {year} {2021})}\BibitemShut {NoStop}%
	\bibitem [{\citenamefont {Korolev}\ \emph
		{et~al.}(2024{\natexlab{a}})\citenamefont {Korolev}, \citenamefont
		{Bashmakova}, \citenamefont {Tagantsev},\ and\ \citenamefont
		{Golubeva}}]{PhysRevA.109.052428}%
	\BibitemOpen
	\bibfield  {author} {\bibinfo {author} {\bibfnamefont {S.~B.}\ \bibnamefont
			{Korolev}}, \bibinfo {author} {\bibfnamefont {E.~N.}\ \bibnamefont
			{Bashmakova}}, \bibinfo {author} {\bibfnamefont {A.~K.}\ \bibnamefont
			{Tagantsev}},\ and\ \bibinfo {author} {\bibfnamefont {T.~Y.}\ \bibnamefont
			{Golubeva}},\ }\bibfield  {title} {\bibinfo {title} {Generation of squeezed
			fock states by measurement},\ }\href
	{https://doi.org/10.1103/PhysRevA.109.052428} {\bibfield  {journal} {\bibinfo
			{journal} {Phys. Rev. A}\ }\textbf {\bibinfo {volume} {109}},\ \bibinfo
		{pages} {052428} (\bibinfo {year} {2024}{\natexlab{a}})}\BibitemShut
	{NoStop}%
	\bibitem [{\citenamefont {Korolev}\ \emph
		{et~al.}(2024{\natexlab{b}})\citenamefont {Korolev}, \citenamefont
		{Bashmakova},\ and\ \citenamefont
		{Golubeva}}]{korolev2024estimationsetstatesobtained}%
	\BibitemOpen
	\bibfield  {author} {\bibinfo {author} {\bibfnamefont {S.~B.}\ \bibnamefont
			{Korolev}}, \bibinfo {author} {\bibfnamefont {E.~N.}\ \bibnamefont
			{Bashmakova}},\ and\ \bibinfo {author} {\bibfnamefont {T.~Y.}\ \bibnamefont
			{Golubeva}},\ }\href {https://arxiv.org/abs/2407.01273} {\bibinfo {title}
		{Estimation of the set of states obtained in particle number measurement
			schemes}} (\bibinfo {year} {2024}{\natexlab{b}}),\ \Eprint
	{https://arxiv.org/abs/2407.01273} {arXiv:2407.01273 [quant-ph]} \BibitemShut
	{NoStop}%
	\bibitem [{\citenamefont {Gradshteyn}\ and\ \citenamefont
		{Ryzhik}(2014)}]{gradshteyn2014table}%
	\BibitemOpen
	\bibfield  {author} {\bibinfo {author} {\bibfnamefont {I.}~\bibnamefont
			{Gradshteyn}}\ and\ \bibinfo {author} {\bibfnamefont {I.}~\bibnamefont
			{Ryzhik}},\ }\href {https://books.google.ru/books?id=F7jiBQAAQBAJ} {\emph
		{\bibinfo {title} {Table of Integrals, Series, and Products}}}\ (\bibinfo
	{publisher} {Elsevier Science},\ \bibinfo {year} {2014})\BibitemShut
	{NoStop}%
	\bibitem [{\citenamefont {Zinatullin}\ \emph {et~al.}(2022)\citenamefont
		{Zinatullin}, \citenamefont {Korolev}, \citenamefont {Manukhova},\ and\
		\citenamefont {Golubeva}}]{PhysRevA.106.032414}%
	\BibitemOpen
	\bibfield  {author} {\bibinfo {author} {\bibfnamefont {E.~R.}\ \bibnamefont
			{Zinatullin}}, \bibinfo {author} {\bibfnamefont {S.~B.}\ \bibnamefont
			{Korolev}}, \bibinfo {author} {\bibfnamefont {A.~D.}\ \bibnamefont
			{Manukhova}},\ and\ \bibinfo {author} {\bibfnamefont {T.~Y.}\ \bibnamefont
			{Golubeva}},\ }\bibfield  {title} {\bibinfo {title} {Error of an arbitrary
			single-mode gaussian transformation on a weighted cluster state using a cubic
			phase gate},\ }\href {https://doi.org/10.1103/PhysRevA.106.032414} {\bibfield
		{journal} {\bibinfo  {journal} {Phys. Rev. A}\ }\textbf {\bibinfo {volume}
			{106}},\ \bibinfo {pages} {032414} (\bibinfo {year} {2022})}\BibitemShut
	{NoStop}%
	\bibitem [{\citenamefont {Braunstein}(2005)}]{PhysRevA.71.055801}%
	\BibitemOpen
	\bibfield  {author} {\bibinfo {author} {\bibfnamefont {S.~L.}\ \bibnamefont
			{Braunstein}},\ }\bibfield  {title} {\bibinfo {title} {Squeezing as an
			irreducible resource},\ }\href {https://doi.org/10.1103/PhysRevA.71.055801}
	{\bibfield  {journal} {\bibinfo  {journal} {Phys. Rev. A}\ }\textbf {\bibinfo
			{volume} {71}},\ \bibinfo {pages} {055801} (\bibinfo {year}
		{2005})}\BibitemShut {NoStop}%
	\bibitem [{\citenamefont {Yoshikawa}\ \emph {et~al.}(2008)\citenamefont
		{Yoshikawa}, \citenamefont {Miwa}, \citenamefont {Huck}, \citenamefont
		{Andersen}, \citenamefont {van Loock},\ and\ \citenamefont
		{Furusawa}}]{PhysRevLett.101.250501}%
	\BibitemOpen
	\bibfield  {author} {\bibinfo {author} {\bibfnamefont {J.-i.}\ \bibnamefont
			{Yoshikawa}}, \bibinfo {author} {\bibfnamefont {Y.}~\bibnamefont {Miwa}},
		\bibinfo {author} {\bibfnamefont {A.}~\bibnamefont {Huck}}, \bibinfo {author}
		{\bibfnamefont {U.~L.}\ \bibnamefont {Andersen}}, \bibinfo {author}
		{\bibfnamefont {P.}~\bibnamefont {van Loock}},\ and\ \bibinfo {author}
		{\bibfnamefont {A.}~\bibnamefont {Furusawa}},\ }\bibfield  {title} {\bibinfo
		{title} {Demonstration of a quantum nondemolition sum gate},\ }\href
	{https://doi.org/10.1103/PhysRevLett.101.250501} {\bibfield  {journal}
		{\bibinfo  {journal} {Phys. Rev. Lett.}\ }\textbf {\bibinfo {volume} {101}},\
		\bibinfo {pages} {250501} (\bibinfo {year} {2008})}\BibitemShut {NoStop}%
	\bibitem [{\citenamefont {Demkowicz-Dobrzański}\ \emph
		{et~al.}(2015)\citenamefont {Demkowicz-Dobrzański}, \citenamefont
		{Jarzyna},\ and\ \citenamefont {Kołodyński}}]{DEMKOWICZDOBRZANSKI2015345}%
	\BibitemOpen
	\bibfield  {author} {\bibinfo {author} {\bibfnamefont {R.}~\bibnamefont
			{Demkowicz-Dobrzański}}, \bibinfo {author} {\bibfnamefont {M.}~\bibnamefont
			{Jarzyna}},\ and\ \bibinfo {author} {\bibfnamefont {J.}~\bibnamefont
			{Kołodyński}},\ }\bibfield  {title} {\bibinfo {title} {Chapter four -
			quantum limits in optical interferometry}\ }(\bibinfo  {publisher}
	{Elsevier},\ \bibinfo {year} {2015})\ pp.\ \bibinfo {pages}
	{345--435}\BibitemShut {NoStop}%
	\bibitem [{\citenamefont {Barnett}\ \emph {et~al.}(1998)\citenamefont
		{Barnett}, \citenamefont {Jeffers}, \citenamefont {Gatti},\ and\
		\citenamefont {Loudon}}]{PhysRevA.57.2134}%
	\BibitemOpen
	\bibfield  {author} {\bibinfo {author} {\bibfnamefont {S.~M.}\ \bibnamefont
			{Barnett}}, \bibinfo {author} {\bibfnamefont {J.}~\bibnamefont {Jeffers}},
		\bibinfo {author} {\bibfnamefont {A.}~\bibnamefont {Gatti}},\ and\ \bibinfo
		{author} {\bibfnamefont {R.}~\bibnamefont {Loudon}},\ }\bibfield  {title}
	{\bibinfo {title} {Quantum optics of lossy beam splitters},\ }\href
	{https://doi.org/10.1103/PhysRevA.57.2134} {\bibfield  {journal} {\bibinfo
			{journal} {Phys. Rev. A}\ }\textbf {\bibinfo {volume} {57}},\ \bibinfo
		{pages} {2134} (\bibinfo {year} {1998})}\BibitemShut {NoStop}%
	\bibitem [{\citenamefont {Oszmaniec}\ and\ \citenamefont
		{Brod}(2018)}]{Oszmaniec_2018}%
	\BibitemOpen
	\bibfield  {author} {\bibinfo {author} {\bibfnamefont {M.}~\bibnamefont
			{Oszmaniec}}\ and\ \bibinfo {author} {\bibfnamefont {D.~J.}\ \bibnamefont
			{Brod}},\ }\bibfield  {title} {\bibinfo {title} {Classical simulation of
			photonic linear optics with lost particles},\ }\href
	{https://doi.org/10.1088/1367-2630/aadfa8} {\bibfield  {journal} {\bibinfo
			{journal} {New Journal of Physics}\ }\textbf {\bibinfo {volume} {20}},\
		\bibinfo {pages} {092002} (\bibinfo {year} {2018})}\BibitemShut {NoStop}%
	\bibitem [{\citenamefont {Qi}\ \emph {et~al.}(2020)\citenamefont {Qi},
		\citenamefont {Brod}, \citenamefont {Quesada},\ and\ \citenamefont
		{Garc\'{\i}a-Patr\'on}}]{PhysRevLett.124.100502}%
	\BibitemOpen
	\bibfield  {author} {\bibinfo {author} {\bibfnamefont {H.}~\bibnamefont
			{Qi}}, \bibinfo {author} {\bibfnamefont {D.~J.}\ \bibnamefont {Brod}},
		\bibinfo {author} {\bibfnamefont {N.}~\bibnamefont {Quesada}},\ and\ \bibinfo
		{author} {\bibfnamefont {R.}~\bibnamefont {Garc\'{\i}a-Patr\'on}},\
	}\bibfield  {title} {\bibinfo {title} {Regimes of classical simulability for
			noisy gaussian boson sampling},\ }\href
	{https://doi.org/10.1103/PhysRevLett.124.100502} {\bibfield  {journal}
		{\bibinfo  {journal} {Phys. Rev. Lett.}\ }\textbf {\bibinfo {volume} {124}},\
		\bibinfo {pages} {100502} (\bibinfo {year} {2020})}\BibitemShut {NoStop}%
	\bibitem [{\citenamefont {Garc{\'{i}}a-Patr{\'{o}}n}\ \emph
		{et~al.}(2019)\citenamefont {Garc{\'{i}}a-Patr{\'{o}}n}, \citenamefont
		{Renema},\ and\ \citenamefont
		{Shchesnovich}}]{GarciaPatron2019simulatingboson}%
	\BibitemOpen
	\bibfield  {author} {\bibinfo {author} {\bibfnamefont {R.}~\bibnamefont
			{Garc{\'{i}}a-Patr{\'{o}}n}}, \bibinfo {author} {\bibfnamefont {J.~J.}\
			\bibnamefont {Renema}},\ and\ \bibinfo {author} {\bibfnamefont
			{V.}~\bibnamefont {Shchesnovich}},\ }\bibfield  {title} {\bibinfo {title}
		{Simulating boson sampling in lossy architectures},\ }\href
	{https://doi.org/10.22331/q-2019-08-05-169} {\bibfield  {journal} {\bibinfo
			{journal} {{Quantum}}\ }\textbf {\bibinfo {volume} {3}},\ \bibinfo {pages}
		{169} (\bibinfo {year} {2019})}\BibitemShut {NoStop}%
	\bibitem [{\citenamefont {Mart{\'{i}}nez-Cifuentes}\ \emph
		{et~al.}(2023)\citenamefont {Mart{\'{i}}nez-Cifuentes}, \citenamefont
		{Fonseca-Romero},\ and\ \citenamefont
		{Quesada}}]{MartinezCifuentes2023classicalmodelsmay}%
	\BibitemOpen
	\bibfield  {author} {\bibinfo {author} {\bibfnamefont {J.}~\bibnamefont
			{Mart{\'{i}}nez-Cifuentes}}, \bibinfo {author} {\bibfnamefont {K.~M.}\
			\bibnamefont {Fonseca-Romero}},\ and\ \bibinfo {author} {\bibfnamefont
			{N.}~\bibnamefont {Quesada}},\ }\bibfield  {title} {\bibinfo {title}
		{Classical models may be a better explanation of the {J}iuzhang 1.0
			{G}aussian {B}oson {S}ampler than its targeted squeezed light model},\ }\href
	{https://doi.org/10.22331/q-2023-08-08-1076} {\bibfield  {journal} {\bibinfo
			{journal} {{Quantum}}\ }\textbf {\bibinfo {volume} {7}},\ \bibinfo {pages}
		{1076} (\bibinfo {year} {2023})}\BibitemShut {NoStop}%
	\bibitem [{\citenamefont {Vahlbruch}\ \emph {et~al.}(2016)\citenamefont
		{Vahlbruch}, \citenamefont {Mehmet}, \citenamefont {Danzmann},\ and\
		\citenamefont {Schnabel}}]{PhysRevLett.117.110801}%
	\BibitemOpen
	\bibfield  {author} {\bibinfo {author} {\bibfnamefont {H.}~\bibnamefont
			{Vahlbruch}}, \bibinfo {author} {\bibfnamefont {M.}~\bibnamefont {Mehmet}},
		\bibinfo {author} {\bibfnamefont {K.}~\bibnamefont {Danzmann}},\ and\
		\bibinfo {author} {\bibfnamefont {R.}~\bibnamefont {Schnabel}},\ }\bibfield
	{title} {\bibinfo {title} {Detection of 15 db squeezed states of light and
			their application for the absolute calibration of photoelectric quantum
			efficiency},\ }\href {https://doi.org/10.1103/PhysRevLett.117.110801}
	{\bibfield  {journal} {\bibinfo  {journal} {Phys. Rev. Lett.}\ }\textbf
		{\bibinfo {volume} {117}},\ \bibinfo {pages} {110801} (\bibinfo {year}
		{2016})}\BibitemShut {NoStop}%
	\bibitem [{\citenamefont {Kilmer}\ and\ \citenamefont
		{Guha}(2019)}]{PhysRevA.99.032302}%
	\BibitemOpen
	\bibfield  {author} {\bibinfo {author} {\bibfnamefont {T.}~\bibnamefont
			{Kilmer}}\ and\ \bibinfo {author} {\bibfnamefont {S.}~\bibnamefont {Guha}},\
	}\bibfield  {title} {\bibinfo {title} {Boosting linear-optical bell
			measurement success probability with predetection squeezing and imperfect
			photon-number-resolving detectors},\ }\href
	{https://doi.org/10.1103/PhysRevA.99.032302} {\bibfield  {journal} {\bibinfo
			{journal} {Phys. Rev. A}\ }\textbf {\bibinfo {volume} {99}},\ \bibinfo
		{pages} {032302} (\bibinfo {year} {2019})}\BibitemShut {NoStop}%
	\bibitem [{\citenamefont {Jozsa}(1994)}]{Jozsa_fidelity}%
	\BibitemOpen
	\bibfield  {author} {\bibinfo {author} {\bibfnamefont {R.}~\bibnamefont
			{Jozsa}},\ }\bibfield  {title} {\bibinfo {title} {Fidelity for mixed quantum
			states},\ }\href {https://doi.org/10.1080/09500349414552171} {\bibfield
		{journal} {\bibinfo  {journal} {Journal of Modern Optics}\ }\textbf {\bibinfo
			{volume} {41}},\ \bibinfo {pages} {2315} (\bibinfo {year} {1994})},\ \Eprint
	{https://arxiv.org/abs/https://doi.org/10.1080/09500349414552171}
	{https://doi.org/10.1080/09500349414552171} \BibitemShut {NoStop}%
	\bibitem [{\citenamefont {Lita}\ \emph {et~al.}(2008)\citenamefont {Lita},
		\citenamefont {Miller},\ and\ \citenamefont {Nam}}]{Lita:08}%
	\BibitemOpen
	\bibfield  {author} {\bibinfo {author} {\bibfnamefont {A.~E.}\ \bibnamefont
			{Lita}}, \bibinfo {author} {\bibfnamefont {A.~J.}\ \bibnamefont {Miller}},\
		and\ \bibinfo {author} {\bibfnamefont {S.~W.}\ \bibnamefont {Nam}},\
	}\bibfield  {title} {\bibinfo {title} {Counting near-infrared single-photons
			with 95\% efficiency},\ }\href {https://doi.org/10.1364/OE.16.003032}
	{\bibfield  {journal} {\bibinfo  {journal} {Opt. Express}\ }\textbf {\bibinfo
			{volume} {16}},\ \bibinfo {pages} {3032} (\bibinfo {year}
		{2008})}\BibitemShut {NoStop}%
	\bibitem [{\citenamefont {Los}\ \emph {et~al.}(2024)\citenamefont {Los},
		\citenamefont {Sidorova}, \citenamefont {Lopez-Rodriguez}, \citenamefont
		{Qualm}, \citenamefont {Chang}, \citenamefont {Steinhauer}, \citenamefont
		{Zwiller},\ and\ \citenamefont {Zadeh}}]{10.1063/5.0204340}%
	\BibitemOpen
	\bibfield  {author} {\bibinfo {author} {\bibfnamefont {J.~W.~N.}\
			\bibnamefont {Los}}, \bibinfo {author} {\bibfnamefont {M.}~\bibnamefont
			{Sidorova}}, \bibinfo {author} {\bibfnamefont {B.}~\bibnamefont
			{Lopez-Rodriguez}}, \bibinfo {author} {\bibfnamefont {P.}~\bibnamefont
			{Qualm}}, \bibinfo {author} {\bibfnamefont {J.}~\bibnamefont {Chang}},
		\bibinfo {author} {\bibfnamefont {S.}~\bibnamefont {Steinhauer}}, \bibinfo
		{author} {\bibfnamefont {V.}~\bibnamefont {Zwiller}},\ and\ \bibinfo {author}
		{\bibfnamefont {I.~E.}\ \bibnamefont {Zadeh}},\ }\bibfield  {title} {\bibinfo
		{title} {{High-performance photon number resolving detectors for 850--950 nm
				wavelength range}},\ }\href {https://doi.org/10.1063/5.0204340} {\bibfield
		{journal} {\bibinfo  {journal} {APL Photonics}\ }\textbf {\bibinfo {volume}
			{9}},\ \bibinfo {pages} {066101} (\bibinfo {year} {2024})},\ \Eprint
	{https://arxiv.org/abs/https://pubs.aip.org/aip/app/article-pdf/doi/10.1063/5.0204340/19995689/066101\_1\_5.0204340.pdf}
	{https://pubs.aip.org/aip/app/article-pdf/doi/10.1063/5.0204340/19995689/066101\_1\_5.0204340.pdf}
	\BibitemShut {NoStop}%
	\bibitem [{\citenamefont {Teich}\ and\ \citenamefont
		{Saleh}(1989)}]{MCTeich_1989}%
	\BibitemOpen
	\bibfield  {author} {\bibinfo {author} {\bibfnamefont {M.~C.}\ \bibnamefont
			{Teich}}\ and\ \bibinfo {author} {\bibfnamefont {B.~E.~A.}\ \bibnamefont
			{Saleh}},\ }\bibfield  {title} {\bibinfo {title} {Squeezed state of light},\
	}\href {https://doi.org/10.1088/0954-8998/1/2/006} {\bibfield  {journal}
		{\bibinfo  {journal} {Quantum Optics: Journal of the European Optical Society
				Part B}\ }\textbf {\bibinfo {volume} {1}},\ \bibinfo {pages} {153} (\bibinfo
		{year} {1989})}\BibitemShut {NoStop}%
	\bibitem [{\citenamefont {Dakna}\ \emph {et~al.}(1997)\citenamefont {Dakna},
		\citenamefont {Anhut}, \citenamefont {Opatrn\'y}, \citenamefont {Kn\"oll},\
		and\ \citenamefont {Welsch}}]{PhysRevA.55.3184}%
	\BibitemOpen
	\bibfield  {author} {\bibinfo {author} {\bibfnamefont {M.}~\bibnamefont
			{Dakna}}, \bibinfo {author} {\bibfnamefont {T.}~\bibnamefont {Anhut}},
		\bibinfo {author} {\bibfnamefont {T.}~\bibnamefont {Opatrn\'y}}, \bibinfo
		{author} {\bibfnamefont {L.}~\bibnamefont {Kn\"oll}},\ and\ \bibinfo {author}
		{\bibfnamefont {D.-G.}\ \bibnamefont {Welsch}},\ }\bibfield  {title}
	{\bibinfo {title} {Generating schr\"odinger-cat-like states by means of
			conditional measurements on a beam splitter},\ }\href
	{https://doi.org/10.1103/PhysRevA.55.3184} {\bibfield  {journal} {\bibinfo
			{journal} {Phys. Rev. A}\ }\textbf {\bibinfo {volume} {55}},\ \bibinfo
		{pages} {3184} (\bibinfo {year} {1997})}\BibitemShut {NoStop}%
	\bibitem [{\citenamefont {Clausen}\ \emph {et~al.}(1999)\citenamefont
		{Clausen}, \citenamefont {Dakna}, \citenamefont {Knoell},\ and\ \citenamefont
		{Welsch}}]{Clausen:1999kb}%
	\BibitemOpen
	\bibfield  {author} {\bibinfo {author} {\bibfnamefont {J.}~\bibnamefont
			{Clausen}}, \bibinfo {author} {\bibfnamefont {M.}~\bibnamefont {Dakna}},
		\bibinfo {author} {\bibfnamefont {L.}~\bibnamefont {Knoell}},\ and\ \bibinfo
		{author} {\bibfnamefont {D.~G.}\ \bibnamefont {Welsch}},\ }\bibfield  {title}
	{\bibinfo {title} {{Conditional quantum state engineering at beam splitter
				arrays}},\ }\href@noop {} {\bibfield  {journal} {\bibinfo  {journal} {Acta
				Phys. Slov.}\ }\textbf {\bibinfo {volume} {49}},\ \bibinfo {pages} {653}
		(\bibinfo {year} {1999})},\ \Eprint {https://arxiv.org/abs/quant-ph/9905085}
	{arXiv:quant-ph/9905085} \BibitemShut {NoStop}%
\end{thebibliography}
\end{document}